\RequirePackage{fix-cm}
\documentclass{svjour3}                     % onecolumn (standard format)
\smartqed  % flush right qed marks, e.g. at end of proof
\usepackage{graphicx}
\usepackage{color}
\usepackage{epsfig}
\usepackage{booktabs}
\usepackage{latexsym}
\usepackage{amsmath}
\usepackage{eufrak}
\usepackage{hyperref}
\usepackage{subfigure}
\usepackage{ulem}
\usepackage{lettrine}
\usepackage{verbatim}
\usepackage{array}
\usepackage{varwidth}
\usepackage{mathptmx} 
\journalname{Flow Turbulence and Combustion}
\begin{document}

\title{An SMLD joint PDF model for turbulent non-premixed combustion using the flamelet progress-variable approach}

\titlerunning{An SMLD joint PDF model for turbulent non-premixed combustion}

\author{Alessandro~Coclite\and Giuseppe~Pascazio\and Pietro~De~Palma\and Luigi~Cutrone\and Matthias~Ihme}

\authorrunning{A. Coclite et al.} 

\institute{Alessandro Coclite \at
Dipartimento di Meccanica, Matematica e Management (DMMM), Politecnico di Bari,\\ 
Via Re David 200 -- 70125 Bari, Italy \\
Centro di Eccellenza in Meccanica Computazionale (CEMeC), Politecnico di Bari,\\ 
Via Re David 200 -- 70125 Bari, Italy\\     
\email{alessandro.coclite@poliba.it}\\
\and
Giuseppe Pascazio \at
Dipartimento di Meccanica, Matematica e Management (DMMM), Politecnico di Bari,\\ 
Via Re David 200 -- 70125 Bari, Italy \\
Centro di Eccellenza in Meccanica Computazionale (CEMeC), Politecnico di Bari,\\ 
Via Re David 200 -- 70125 Bari, Italy\\     
\email{giuseppe.pascazio@poliba.it}\\
\and
Pietro De Palma \at
Dipartimento di Meccanica, Matematica e Management (DMMM), Politecnico di Bari,\\ 
Via Re David 200 -- 70125 Bari, Italy \\
Centro di Eccellenza in Meccanica Computazionale (CEMeC), Politecnico di Bari,\\ 
Via Re David 200 -- 70125 Bari, Italy\\     
\email{pietro.depalma@poliba.it}\\
\and
Luigi Cutrone \at
Centro Italiano Ricerche Aerospaziali (CIRA),\\ 
Via Maiorise -- 81043 Capua, Italy \\
Centro di Eccellenza in Meccanica Computazionale (CEMeC), Politecnico di Bari,\\ 
Via Re David 200 -- 70125 Bari, Italy\\     
\email{l.cutrone@cira.it}\\
\and 
Matthias Ihme \at
Department of Mechanical Engineering, Stanford University,\\ 
Stanford, CA 94305, USA\\   
\email{mihme@stanford.edu}\\
}

\date{Received: date / Accepted: date% The correct dates will be entered by the editor
}

\maketitle

\begin{abstract}
This paper provides an improved flamelet/progress variable (FPV) model for the simulation of turbulent combustion, employing the statistically most likely distribution (SMLD) approach for the joint probability density function (PDF) of the mixture fraction, $Z$, and of the progress parameter, $\Lambda$. Steady-state FPV models are built presuming the functional shape of the joint PDF of $Z$ and $\Lambda$ in order to evaluate Favre-averages of thermodynamic quantities. The mixture fraction is widely assumed to behave as a passive scalar with a mono-modal behaviour modelled by a $\beta$-distribution. Moreover, under the hypothesis that $Z$ and $\Lambda$ are statistically independent, the joint PDF coincides with the product of the two marginal PDFs. In this work we discuss these two constitutive hypotheses. The proposed model evaluates the most probable joint distribution of $Z$ and $\Lambda$, relaxing some crucial assumption on their statistical behaviour. This provides a more general model in the context of FPV approach and an effective tool to verify the adequateness of widely used hypotheses. The model is validated versus experimental data of well-known test cases, namely, the Sandia flames. The results are also compared with those obtained by the standard FPV approach, analysing the role of the PDF functional form on turbulent combustion simulations. 

\keywords{Presumed PDF \and Methane combustion \and Turbulent non-premixed combustion \and Reynolds-Averaged Navier--Stokes equations}
\end{abstract}

\section{Introduction}
\label{intro}

\begin{figure}
\begin{center}
\includegraphics[scale=0.2]{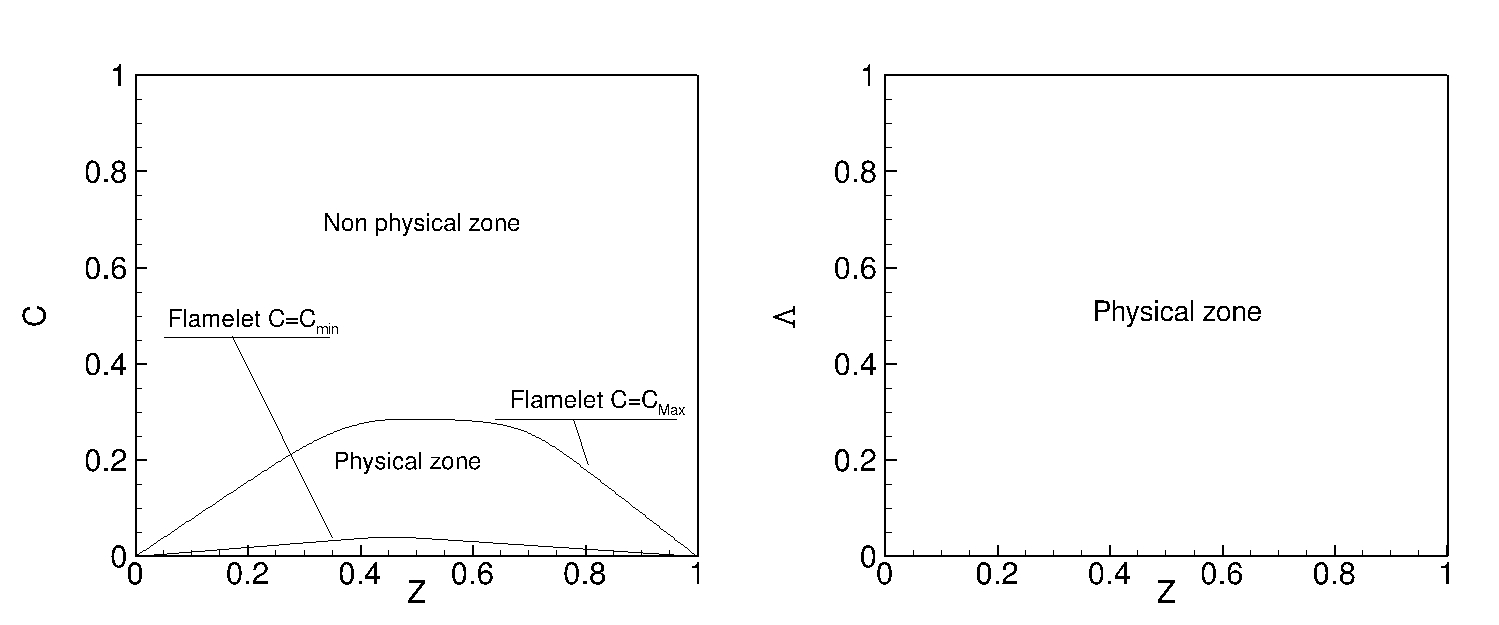}
\caption{Schematic of the procedure for the conditional normalization of the progress variable.}
\label{CZ_Norm}
\end{center}
\end{figure}

\begin{figure}
\begin{center}
\includegraphics[scale=0.1]{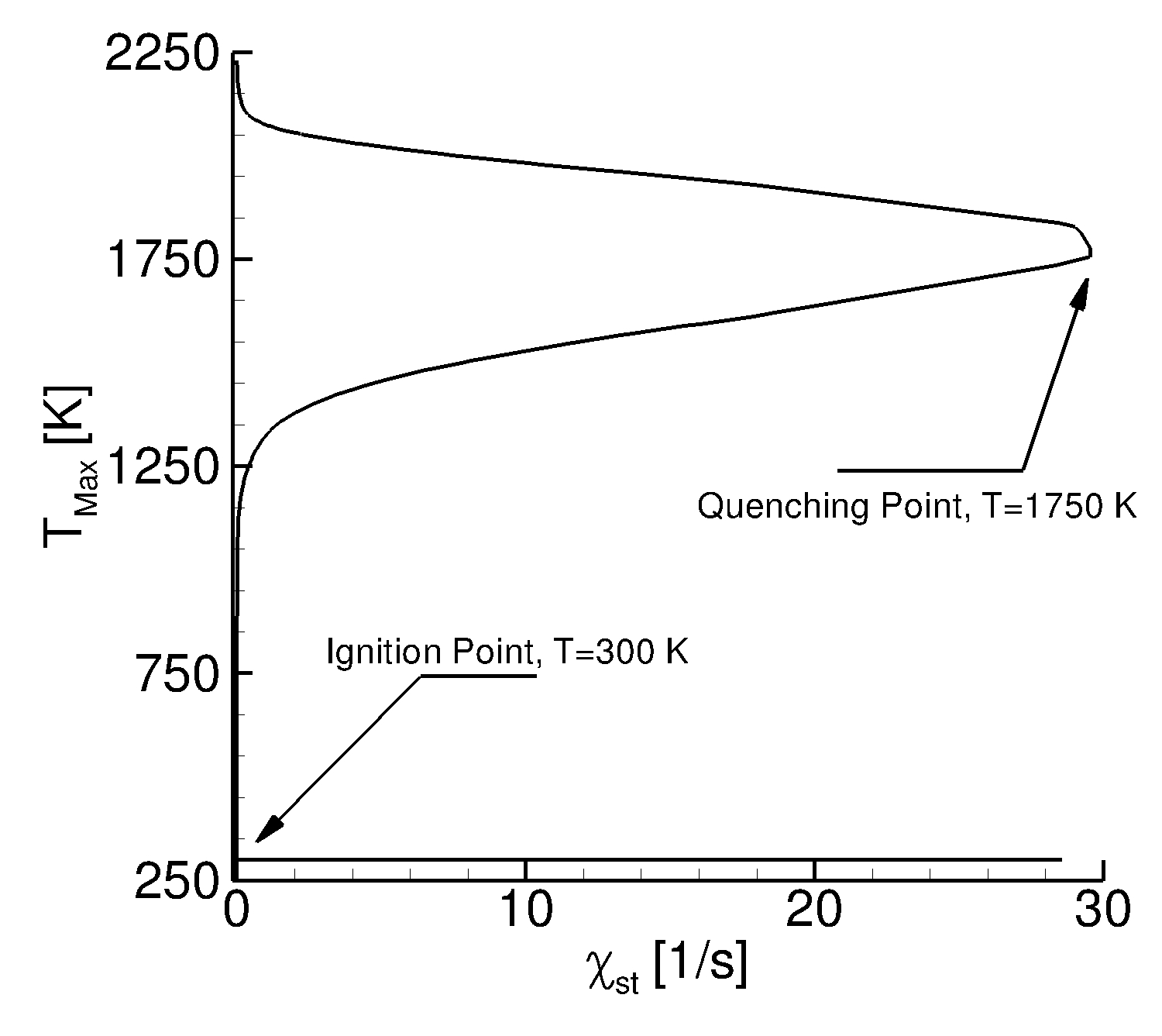}
\caption{Distribution obtained plotting the maximum temperature of each flamelet versus the scalar stoichiometric dissipation rate of about 250 flamelet with locally refined distribution in $\chi_{st}$ direction.}
\label{scurve}
\end{center}
\end{figure}

The development of new technologies to enhance and control combustion processes is nowadays fundamental in order to increase efficiency and reduce emissions 
in many engineering applications, such as internal combustion engines, advanced gas turbine systems, pulse detonation engines, high-speed air-breathing propulsion devices. 
In this context, the simulation of turbulent reacting flows is very useful to reduce experimental costs and to advance the comprehension of the basic physical mechanisms. 
Turbulent combustion is a multi-scale problem, where the interaction between chemical kinetics, molecular, and turbulent transport occurs over a wide range of length and time scales. 
The numerical simulation of such phenomena with detailed chemistry is today still prohibitive. The huge computational cost stems from the fact that,
even for a simple fuels, detailed kinetic mechanism can involve thousands of reactions and, consequently, hundreds of chemical species.  
One way to overcome this problem is to simplify it using
reduced chemical models with a limited number of reactions and species involved, but still the number of degrees of freedom may
be prohibitive for applications of engineering interest.
Therefore, several approaches have been proposed that simplify the problem by pre-computing,
in terms of a small number of variables, the entire chemical state-space \cite{maas,piercemoin2004,laminarhydrogen,oijen}. 
Among such models, the flamelet-progress variable
(FPV) approach~\cite{piercemoin2004,ihmeal2005} is considered in the present work.
For the case of non-premixed combustion of interest here, the basic assumptions of the flamelet model are fulfilled
for sufficiently large Damk\"{o}hler number, $Da$. In fact, when the reaction zone thickness is very thin with respect to the Kolmogorov
length scale, turbulent structures are unable to penetrate into the reaction zone and cannot destroy the laminar flame structure. 
Effects of turbulence only result in a deformation and
straining of the flame sheet and locally the flame structure can be described as function of the mixture fraction,  $Z$,
the scalar dissipation rate, $\chi$, and the time. The scalar dissipation rate, $\chi=2D_Z(\nabla Z)^2$, is a measure of the gradient of the mixture fraction 
representing the molecular diffusion of the species in the flame region, $D_Z$ being the molecular diffusion coefficient.
Therefore, the entire flame behaviour can be obtained as a combination of solutions of the laminar flamelet equations.
In the present work we consider a further simplification assuming a steady flamelet behaviour, so that chemical effects are entirely
determined by the value of $Z$. Moreover, $\chi$ defines the effects of the flow on the flame structure according to
the following steady laminar flamelet equation (SLFE)~\cite{peters} for the generic variable $\phi$
\begin{equation}
\label{slfe}
{-\rho\frac{\chi}{2}\frac{\partial^2 \phi}{\partial Z^2}=\dot\omega_\phi}\, ,
\end{equation} 
where $\rho$ is the density and  $\dot\omega_\phi$ is the source term related to $\phi$~\cite{piercemoin2004}, different from zero in the case of finite rate chemistry. 
In this work, we employ the FPV model proposed by Pierce and Moin~\cite{pierce,piercemoin2004} to evaluate all of the thermo-chemical quantities involved in the combustion process. This approach is based on the parametrization of the generic thermo-chemical quantity, $\phi$, in terms of the mixture fraction, $Z$, and of the progress parameter, $\Lambda$, instead of $\chi$:
\begin{equation}
\label{phi}
{\phi=F_\phi(Z,\Lambda)}\, .
\end{equation}
Using such a parameter, independent of the mixture fraction, one can uniquely identify each flame state along the stable and unstable branches of the S-shaped curve. A suitable definition of $\Lambda$ leads to a substantial simplification of the presumed PDF-closure model. In contrast, the solution of the transport equation for $\Lambda$ is quite complex since it requires non-trivial modelling of several unclosed terms~\cite{ihmea}. In order to overcome such a difficulty, the progress parameter is derived from a reaction progress variable, $C$, such as the temperature or a linear combination of the main reaction products, whose behaviour is governed by a simpler transport equation. Therefore, a transport equation for $C$ is solved and the flamelet library is parametrized in terms of $Z$ and $C$. Requiring that the transformation between $\Lambda$ and $C$ be bijective, from equation~\eqref{slfe} one has
\begin{equation}
\label{lambdaFuncC}
\Lambda=F^{-1}_C(Z,C)\, ,
\end{equation} 
and any thermo-chemical variable can then be expressed as:
\begin{equation}
\phi=F_\phi(Z,F^{-1}_C(Z,C))\, .
\end{equation}
The choice of the progress variable is not unique and some recent works discuss in details this issue proposing a procedure for the selection of the optimal definition of $C$~\cite{ihmejcp2012,cuenotCF2012,vervischCF2013}. Here the progress variable is defined by the sum of the mass fractions of the main products; for methane combustion:
\begin{equation}
\label{progress_variable}
{C=Y_{H_2}+Y_{H_2O}+Y_{CO}+Y_{CO_2}}\, .
\end{equation} 
This turns out to be a suitable choice for the Sandia flames test case considered in this work~\cite{ihmejcp2012}. 
Observe that, since only the main products of a combustion process are considered the maximum value assumed by the progress variable is smaller than unity. A stretching with respect to the minimum and maximum conditioned value of $C$ over the mixture fraction is made, namely:
\begin{equation}
\label{normalization}
\Lambda = \frac{{C} - {C_{Min}|Z}}{{C_{Max}|Z} - {C_{Min}|Z}}\, .
\end{equation}
This definition greatly simplifies the presumed PDF closure procedure, recovering the unitary length of the variation interval for the progress parameter~\cite{klimenko99} and so overcoming the issue in the integrals evaluation. 
Figure~\ref{CZ_Norm} shows the effect of such a normalization.
Equation~\eqref{phi} is taken as the solution of the SLFE~\eqref{slfe}. Each solution corresponding to a given value of $\chi$ is a flamelet and the
solution variety over $\chi = \chi_{st}$ is shown in figure~\ref{scurve}.
From equation \eqref{phi} one can obtain the Favre-averaged value of $\phi$ and of its variance using the definitions:
\begin{equation}
\label{media}
{\widetilde\phi=\int\int F_\phi(Z,\Lambda)\widetilde{P}(Z,\Lambda)dZ d\Lambda},
\end{equation}
\begin{equation}
\label{varianza}
{\widetilde{\phi''^2}=\int\int [F_\phi(Z,\Lambda)-\widetilde\phi]^2\ \widetilde{P}(Z,\Lambda)dZd\Lambda},
\end{equation}
where $\widetilde P(Z,\Lambda)$ is the density-weighted PDF.
This function plays a crucial role in the definition of the model, affecting both its accuracy and computational cost.
Moreover, the choice of $\widetilde P(Z,\Lambda)$ is not straightforward because of the unknown statistical behaviour of the two variables $Z$ and $\Lambda$~\cite{ihmea} and its definition is still an open problem whose solution is being pursued by several researches~\cite{ihmeal2005,ihmea,ihmeb,DeMeesterCF2012,AbrahamPoF2012}.

The aim of this work is to provide an analysis of state-of-the-art FPV models and to propose a more general scheme based on the statistically most likely distribution (SMLD)~\cite{pope} approach for the joint PDF of $Z$ and $\Lambda$. In fact, this allows one to avoid any assumption about the statistical correlation of the two variables and about the behaviour of the mixture fraction. In  order to assess the capability of the combustion models, they are employed here in conjunction with a Reynolds-Averaged Navier--Stokes (RANS) solver with a $k$-$\omega$ turbulence model closure. 

This work is organised as follow: in section~\ref{intro} a brief introduction of the state-of-the-art is presented in order to show the motivation and the basic outlines of \textit{flamelet} models; then, in sections~\ref{fpv} and~\ref{goveq} the theoretical and numerical details of the models employed are provided. The comparison between four models for the definition of the presumed-PDF is discussed in section~\ref{results} and numerical results obtained studying two subsonic flames, namely Sandia Laboratories flame~D~and~E, are presented. Finally, some conclusions are provided.

\section{The flamelet progress variable models}
\label{fpv}

\subsection{The standard FPV model (model A)}

In this section, the standard FPV model (called here model A) is briefly described. It is based on the following hypotheses (Hp).\\
$Hp\ 1$: The mixture fraction $Z$ and the progress parameter $\Lambda$ are the two independent variables of the combustion model; 
they form a basis by which one can derive all of the thermo-chemical quantities.\\
$Hp\ 2$: The steady flamelet assumption is used, so that the solution variety of equation~\eqref{slfe} is made over $\chi=\chi_{st}$, see figure~\ref{scurve}. 
A key difficulty to integrate the flamelet equations is to know a priori information about the scalar dissipation rate dependence on the mixture fraction $\chi(Z)$. In fact, flamelet libraries are computed in advance and are assumed to be independent of the flow field. Therefore, the dependence of $\chi$ on $Z$ has to be modelled~\cite{pitsch98,pitschchenpeters98,kimwilliams93}. In this work, the functional form of $\chi(Z)$ has been taken from an idealized configuration. Peters~\cite{peters84} developed an analytic expression for the distribution of the scalar dissipation rate in a counterflow diffusion flame, which leads to an inverse error functional:
\begin{equation}
\label{chi_st}
{\chi(Z)=\chi_{st}\frac{\Phi(Z)}{\Phi(Z_{st})}},
\end{equation}
where $Z_{st}$ is the mixture fraction evaluated at the stoichiometric point and $\Phi(Z)$ is the distribution of scalar dissipation rate for the counterflow diffusion flame~\cite{peters84}. \\
$Hp\ 3$: The PDF in \eqref{media} and \eqref{varianza} is presumed and its choice establishes the statistical correlation between $Z$ and $\Lambda$. Therefore, employing Bayes' theorem, 
\begin{equation}
\label{bayes}
%%%%%{\widetilde{P}(Z,C)=\widetilde P(Z)\widetilde P(\Lambda|Z)} \, ,
{\widetilde{P}(Z,\Lambda)=\widetilde P(Z)P(\Lambda|Z)} \, ,
\end{equation}
to evaluate such a PDF, one usually presumes the functional shape of the marginal PDF of $Z$ and of the conditional PDF of $\Lambda|Z$. \\
$Hp\ 4$: The statistical behaviour of the mixture fraction is described by a $\beta$-distribution.
In fact, even though the definition of $\widetilde P(Z)$ is still an open question~\cite{pope}, 
it has been shown by several authors that the mixture fraction behaves like a conserved scalar whose statistical distribution
can be approximated by a $\beta$-function~\cite{cook,jimenez,wall}. 
The two parameter family of the $\beta$-distribution in the interval $x\in [0,1]$ is given by:
\begin{equation}
\label{beta}
{\beta(x;\widetilde{x},\widetilde{x''^2})=x^{a-1}(1-x)^{b-1}\frac{\Gamma(a+b)}{\Gamma(a)\Gamma(b)}},
\end{equation}
where $\Gamma(x)$ is the Gamma-function and $a$ and $b$ are two parameters related to $\widetilde x$ and $\widetilde{x''^2}$
\begin{equation}
\label{aeb}
{a=\frac{\widetilde x(\widetilde x- \widetilde{x}^2-\widetilde{x''^2})}{\widetilde{x''^2}}, \ \ \ b=\frac{(1-\widetilde x)(\widetilde x-\widetilde{x}^2-\widetilde{x''^2})}{\widetilde{x''^2}}}.
\end{equation}\\
$Hp\ 5$: Statistical independence of $Z$ and $\Lambda$ is assumed, so that $\widetilde P(Z,\Lambda)=\widetilde P(Z)P(\Lambda)$.\\
$Hp\ 6$: $P(\Lambda)$ is a Dirac distribution, implying a great simplification in the theoretical framework.

Under these assumptions, it can be shown that there is only one solution of equation \eqref{slfe} for each value of the scalar dissipation rate. With these criteria, the Favre-average of a generic thermo-chemical quantity is given by:
\begin{equation}
\label{phidelta}
{\widetilde\phi=\int\int F_\phi(Z,\Lambda)\widetilde \beta(Z)\delta(\Lambda-\widetilde{\Lambda})dZd\Lambda=\int F_\phi(Z,\widetilde \Lambda)\widetilde \beta(Z)dZ}\, .
\end{equation} 
Therefore, one has only three additional transport equations (for $\widetilde Z$, $\widetilde{Z''^2}$ and $\widetilde C$) to evaluate all of the thermo-chemical quantities in the flow, thus avoiding the expensive solution of one transport equation for each chemical species. 

\subsection{The FPV model with $\beta$-distribution for $P(\Lambda)$ (model B)}

It is well known that a reactive scalar, such as $C$, depends on a combination of solutions of equation~\eqref{slfe} for each chemical state and therefore 
its PDF cannot be accurately approximated by a Dirac distribution~\cite{ihmea}.
Thus, a different model (called here model~B) has been proposed in~\cite{cha_pitsch2002},
in which $Z$ and $\Lambda$ behave in the same way, namely, following a $\beta$-distribution:
\begin{equation}
\label{betabeta}
{\widetilde P(Z,\Lambda)=\widetilde\beta(Z)\beta(\Lambda)}.
\end{equation}
This reformulation of $Hp\ 6$ can be considered as an improvement with respect to model~A. However, this assumption has an additional cost since model~B requires the evaluation of an additional transport equation for $\widetilde{C''^2}$. 

\subsection{The FPV model with SMLD for $P(\Lambda)$ (model C)}

The probability distribution of a reacting scalar is often multi-modal and its functional form depends on the turbulence-chemistry interaction. 
It is noteworthy that the $\beta$-function is suitable to describe the behaviour of a passive scalar, such as $Z$; 
in fact, it becomes bi-modal for sufficiently large values of the variance. In order to have an improved evaluation of $P(\Lambda)$, a more general distribution that allows  multi-modal behaviour even for a small variance is needed.
Therefore, Ihme and Pitsch~\cite{ihmea} proposed a new model evaluating the conditional PDF as the statistically most likely distribution (SMLD) based on the values of $\widetilde Z$, $\widetilde{Z''^2}$, $\widetilde \Lambda$ and $\widetilde{\Lambda''^2}$.
The rational behind the SMLD approach is that one can construct a PDF based on the knowledge of a small number of moments and on the minimization of the uncertainty in order to have the most probable distribution. In this perspective, model~C is more general than model~A and model~B, since the functional shape of $P(\Lambda)$ is not fixed. 
Relying on the knowledge of the first three moments of $P(\Lambda)$, a unique measure, $S$, of the predictability of a thermodynamic 
state can be defined~\cite{pope,heinz}. 
$S$ is an entropy function depending on $P(\Lambda)$, $S=S(P(\Lambda))$~\cite{shannon}, that can be thought of as the Boltzmann's entropy:
\begin{equation}
\label{s}
{S=-\int P(\Lambda) \ln\Bigl(\frac{P(\Lambda)}{Q(\Lambda)}\Bigr) d\Lambda},
\end{equation}
where $Q(\Lambda)$ is a density function proposed by Pope~\cite{pope} that takes into account the bias in the composition space when no information about the moments are given,
\begin{equation}
\label{pope_q}
{Q(\Lambda)=\frac{1}{\Lambda-\widetilde{\Lambda}}\Bigl[1+\Bigl(\frac{\dot{\omega}_C (\Lambda-\widetilde{\Lambda})}{\widetilde{\chi}_C}\Bigr)^2\Bigr]^{-1/2}}\, .
\end{equation} 
The goal is to construct a PDF that maximizes the entropy $S$. Following the Lagrangian optimization approach, the functional $S^*$ is defined by involving the constraints on the moments:
\begin{equation}
\label{s_star}
{S^*=-\int d\Lambda\Bigl\{  \frac{P(\Lambda)}{Q(\Lambda)} \ln\Bigl(\frac{ P(\Lambda)}{Q(\Lambda)}\Bigr)+\sum_{n=1}^2 \mu_n \Lambda^n \frac{P(\Lambda)}{Q(\Lambda)}-\frac{P(\Lambda)}{Q(\Lambda)} \Bigr\}}.
\end{equation}
In the above equation $\mu_n$ are the Lagrange's multipliers while the last fraction term is introduced to normalize $P(\Lambda)$.
Equation~\eqref{s_star} can be obtained from equation~\eqref{s} independently of the particular form of $Q(\Lambda)$ considering the analytical form of the mean and the variance as constraints. The term depending on $\mu_n$ corresponds to the two constraints given by the first and the second moments of the PDF, namely $\widetilde{\Lambda}$ and $\widetilde{\Lambda''^2}$. The first moment is enforced considering the last term, $\frac{P(\Lambda)}{Q(\Lambda)}$, needed to ensure the unity sum of $P(\Lambda)$.   
The expression for $P(\Lambda)$, obtained by maximising $S^*$, reads: 
\begin{equation}
\label{smld}
{P(\Lambda)=Q(\Lambda)\frac{1}{\mu_0}\exp\Bigl\{-\sum_{n=1}^2 \frac{\mu_n}{n}(\Lambda-\widetilde \Lambda)^n \Bigr\}}\, ,
\end{equation}
where:
\begin{eqnarray}
\label{mu0_C}
\mu_0&=&\int_0^1 d\Lambda P(\Lambda),\\
\label{mu1_C}
-\mu_1&=&\int_0^1 d\Lambda \partial_\Lambda (P(\Lambda))=P(1)-P(0),\\
\label{mu2_C}
1-\mu_2\widetilde{\Lambda''^2}&=&\int_0^1 d\Lambda \partial_\Lambda[(\Lambda-\widetilde \Lambda) P(\Lambda)]=P(1)-\widetilde \Lambda\mu_1,
\end{eqnarray}
since $\Lambda$ is bounded in $[0,1]$.
The model still needs an additional assumption for closure. 
Here, we make a different choice with respect to the original SMLD model of Ihme and Pitsch~\cite{ihmea}.
We assume that the first and the last point of $P(\Lambda)$ are equal to the first and last points of $\beta(\Lambda)$ evaluated with the same 
values of the mean and variance:
\begin{equation}
\label{extrema}
{ P(1;\widetilde \Lambda,\widetilde{\Lambda''^2})=\beta(1;\widetilde \Lambda,\widetilde{\Lambda''^2})},\\
{\ \ \ P(0;\widetilde \Lambda,\widetilde{\Lambda''^2})=\beta(0;\widetilde \Lambda,\widetilde{\Lambda''^2})}.
\end{equation}
This assumption does not affect the multi-modal nature of the distribution, but simplifies the model implementation since
there is no need to evaluate the roots of a non-linear system at each computational point as in the original version of the SMLD model~\cite{ihmea}.
\begin{figure}
\begin{center}
\includegraphics[scale=0.2]{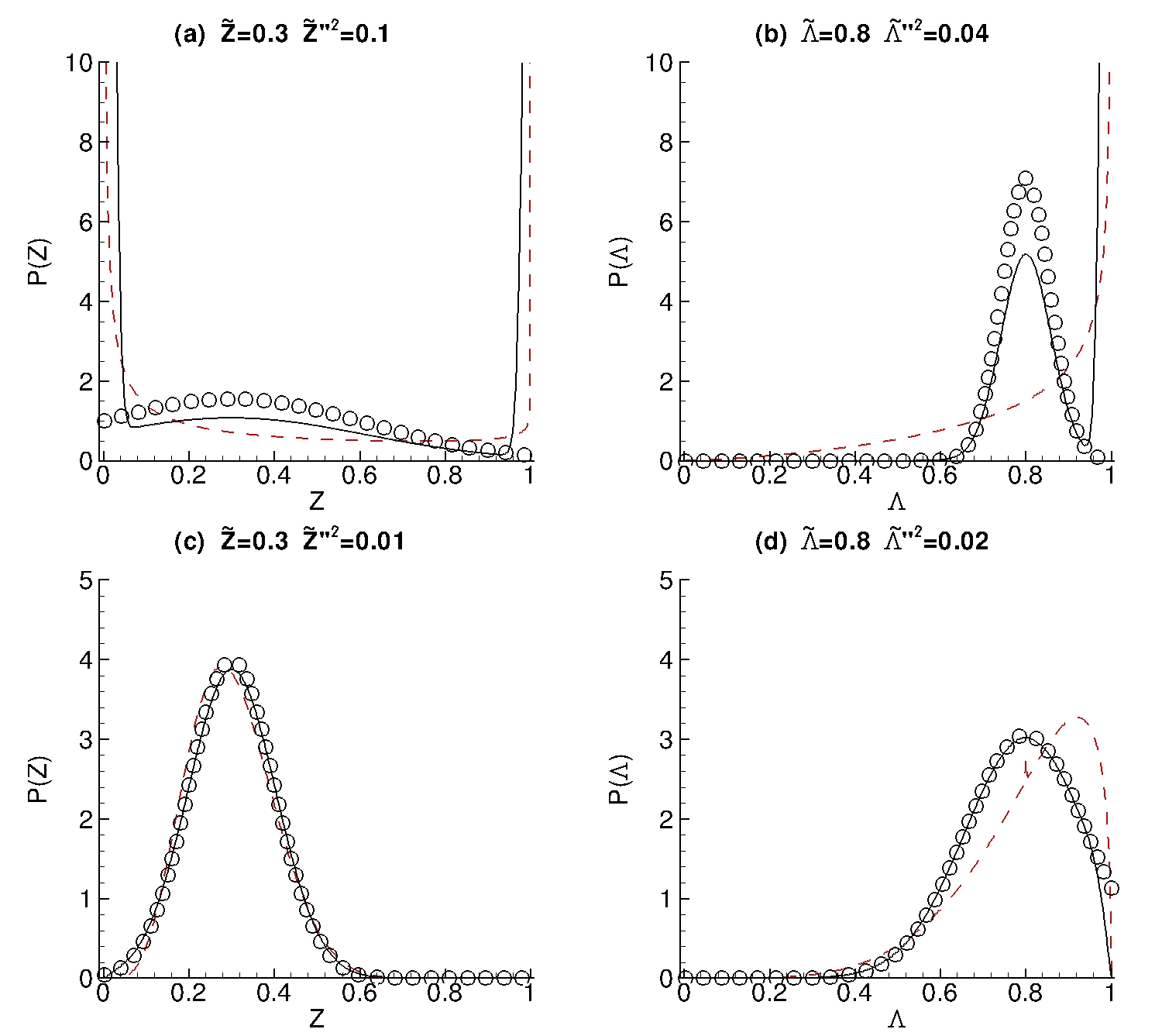}
\caption{Comparison among three PDFs for two set of values of the mean and the variance: $\beta$ distribution (red dashed line); $P_{SML,2}$ (black solid line) and $P_{SML,2}^{Nat}$~\cite{ihmea} (symbols). Boundary values: Panel~a, $P_{SML,2}(0)=\beta(0)=40$, $P_{SML,2}(1)=\beta(1)=17$, $P_{SML,2}^{Nat}(0)=1$, $P_{SML,2}^{Nat}(1)=0$; Panel~b, $P_{SML,2}(0)=\beta(0)=0$, $P_{SML,2}(1)=\beta(1)=13$, $P_{SML,2}^{Nat}(0)=0$, $P_{SML,2}^{Nat}(1)=0$; Panel~c, $P_{SML,2}(0)=\beta(0)=0$, $P_{SML,2}(1)=\beta(1)=0$, $P_{SML,2}^{Nat}(0)=0$, $P_{SML,2}^{Nat}(1)=0$; Panel~d, $P_{SML,2}(0)=\beta(0)=0$, $P_{SML,2}(1)=\beta(1)=0$, $P_{SML,2}^{Nat}(0)=0$, $P_{SML,2}^{Nat}(1)=1.13$.}
\label{P_confronto}
\end{center}
\end{figure}
Figure~\ref{P_confronto} shows the comparison between $P_{SML,2}$, evaluated with the constraint given by equation~\eqref{extrema}~(model~C), and the distribution obtained by solving the natural constraint given by the non-linear system $\delta_{ij}-\mu_{2,ik}\sigma_{kj}=0$~\cite{ihmea}, where $\sigma_{kj}$ is the covariance matrix, namely $P_{SML,2}^{Nat}$.
The two distributions are practically coincident for low and high values of the variance. On the other hand, for its intermediate values, using the constraints in equation~\eqref{extrema} provides a different behaviour of the two PDFs at the extrema of the range of variation of $Z$ and $\Lambda$. However, such a discrepancy is localized at the extrema and only slightly affects the distribution for intermediate values, so that the simplified constraints in equation~\eqref{extrema} can be considered satisfactory.  

\subsection{The FPV model with SMLD for $\widetilde P(Z, \Lambda)$ (model D)}

The major advantage of the SMLD approach over usually employed presumed PDF closure models is that it provides a systematic framework to incorporate an arbitrary number of moment information.
It has been shown~\cite{ihmeal2005} that, even though equation~\eqref{slfe} is based on assuming as independent variables $Z$ and $\Lambda$, one has to properly take into account the statistical correlation between $Z$ and $\Lambda$ due to the effect of turbulence modelling, employing a suitable estimation of the joint PDF. Therefore, we propose a different model (model~D) in which the SMLD approach is applied directly to the joint distribution. 
In this way one does not need hypotheses $4$-$6$ and the most probable PDF can be evaluated considering only the known informations about the statistical means of the joint distribution.
Assuming that the first three moments of the joint probability density function $\widetilde P(\vec x)$, where $\vec x=(Z,\Lambda)^T$, are known,
the procedure employed for model~C is here extended to the case of the two dimensional PDF, obtaining:
\begin{multline}
\label{smld2}
\widetilde P_{SML,2}(Z,\Lambda)= \frac{Q(\Lambda)}{\mu_0}\exp\Bigl\{-\Bigl[\mu_{1,1} (Z-\widetilde Z)+\mu_{1,2}(\Lambda-\widetilde \Lambda)\Bigr]\\-\frac{1}{2}\Bigl[\mu_{2,11}(Z-\widetilde Z)^2+\mu_{2,12}(Z-\widetilde Z)(\Lambda-\widetilde \Lambda)\\+\mu_{2,21}(\Lambda-\widetilde \Lambda)(Z-\widetilde Z)+\mu_{2,22}(\Lambda-\widetilde \Lambda)^2 \Bigl]
\Bigr\}\, .
\end{multline}
In the equation above, $Q(\Lambda|Z)=Q(\Lambda)$ is given by equation~\eqref{pope_q}, being the bias PDF for $Z$ a constant (assumed equal to one), as discussed by Pope~\cite{pope1979} due to the linearity of the mixture fraction transport equation in $Z$.
Moreover , $\mu_0$ is a scalar, $\vec{\mu_1}$ is a two~-~component vector,
and $\overleftrightarrow{\mu_2}$ is a square matrix of rank two:
\begin{eqnarray}
\label{moltiplicatori}
\mu_0&=&\int d\vec x \widetilde P_{SML,2}(\vec x) ,\\
\label{moltiplicatori1}
-\mu_{1,i}&=&\int d\vec x\partial_{x_i} \widetilde P_{SML,2}(\vec x)\nonumber\\&=&\beta(1;\widetilde\xi_i,\widetilde{\xi_i''^2})-\beta(0;\widetilde \xi_i,\widetilde{\xi_i''^2}) ,\\
\label{moltiplicatori2}
\delta_{kl}-\mu_{2,kn}\ \widetilde{\xi'_n\xi'_l}&=&\int d\vec x \partial_{x_k}((x_l-\widetilde\xi_l)\widetilde P_{SML,2}(\vec x))\nonumber\\&=&\beta(1;\widetilde \xi_k,\widetilde{\xi'_k\xi'_l})-\widetilde\xi_k\mu_{1,l} ,
\end{eqnarray} 
where $i$, $k$, $n$, and $l$ indicate the vector components; 
$\widetilde\xi_i$, $\widetilde\xi'_i$ and $\widetilde{\xi^{''2}_i}$ are the mean, the fluctuation and the variance of the $i$-th component of $\vec x$, respectively $\widetilde{x_i}$, $x_i-\widetilde{x_i}$, and $\widetilde{x_i''^2}$.\\
To proof equations \eqref{moltiplicatori}, \eqref{moltiplicatori1} and \eqref{moltiplicatori2} one needs to consider Bayes' theorem applied to $\widetilde P(\vec x)$:
\begin{equation}
{\widetilde P_{SML,2}(\vec x)=\widetilde P_{SML,2}(x_1) P_{SML,2}(x_2|x_1)},
\end{equation}
 where $\widetilde P_{SML,2}(x_1)$ is the marginal PDF and $ P_{SML,2}(x_2|x_1)$ is the conditional one. These two functions are both PDFs, therefore by definition one has:
\begin{equation}
{\int dx_1\widetilde  P_{SML,2}(x_1)=1},
\end{equation} 
\begin{equation}
{\int dx_2 P_{SML,2}(x_2|x_1)=1}.
\end{equation} 
Considering a probability space $(\Omega,F,P)$, where $\Omega$ is the sample space, $F$ the events algebra and $P$ the PDF, and two events $A,B\in \Omega$ with $P(A)\neq 0)$ the conditional probability of $B$ at a given $A$ is: $P(B|A)$. $P(\cdot|A):F\to [0,1]$ is actually a PDF as proved by invoking the characterization theorem, so that, $P(\Omega |A)=1$.
%%To prove that $P(\cdot|A):F\to [0,1]$ is a PDF one can invoke the characterization theorem and show that $P(\cdot|A)$ verify all of the following prepositions~\cite{ross}:
%%\begin{eqnarray}
%%P(\emptyset |A)&=&0,\\
%%P(\Omega |A)&=&P(A|A)=1,\\
%%P(B_1\cup B_2|A)&=&P(B_1|A)+P(B_2|A),\ \ if\ \  B_1\cap B_2=\emptyset,
%%\end{eqnarray}
Therefore, $(\Omega,F,P(\cdot |A))$ is a new probability space in which there is the information that $A$ is an event always verified. In some way one can think that $A$ is a sort of parameter that measures the known region of $\Omega$.\\
Let us consider equation \eqref{moltiplicatori1}:
\begin{eqnarray}
\label{proof}
%%%%%-\mu_{1,1}=\int d\vec x\partial_{x_1} \widetilde P_{SML,2}(\vec x)=\\=\int_0^1 dx_2 \int_0^1 dx_1 \partial_{x_1}(\widetilde P_{SML,2}(x_1) P_{SML,2}(x_2|x_1))= \\ =\int_0^1 dx_2( P_{SML,2}(x_2|x_1=1) \widetilde P_{SML,2}(x_1=1)-\\- P_{SML,2}(x_2|x_1=0)\widetilde P_{SML,2}(x_1=0))=\\=\widetilde P_{SML,2}(x_{1}=1)\int_{0}^{1}dx_{2} P_{SML,2}(x_{2}|x_{1}=1)-\\-\widetilde P_{SML,2}(x_{1}=0)\int_{0}^{1}dx_{2} P_{SML,2}(x_{2}|x_{1}=0)=\\=\widetilde P_{SML,2}(x_1)\mid_0^1 .
%%%%%
-\mu_{1,1}&=&\int d\vec x\partial_{x_1} \widetilde P_{SML,2}(\vec x)=\\
\nonumber
&=&\int_0^1 dx_2 \int_0^1 dx_1 \partial_{x_1}(\widetilde P_{SML,2}(x_1) P_{SML,2}(x_2|x_1))= \\ 
\nonumber
&=&\widetilde P_{SML,2}(x_1)\mid_0^1 .
\end{eqnarray}
Straightforwardly comes the proof of the equations~\eqref{moltiplicatori2} and trivially that of the equation~\eqref{moltiplicatori}.\\
It is noteworthy that, assuming the statistical independence of $Z$ and $\Lambda$ and a $\beta$-distribution for $\widetilde{P}(Z)$, model~D automatically reduces to model~C. Indeed, in such a case, 
\begin{equation}
{\widetilde P_{SML,2}(Z,\Lambda)= \widetilde P(Z) P_{SML,2}(\Lambda)=\widetilde\beta(Z) P_{SML,2}(\Lambda)} \, ,
\end{equation}
where the first multiplier $\mu_0$ is still given by equation \eqref{moltiplicatori}, while the second and the third ones, $\vec{\mu_1}$ and $\overleftrightarrow{\mu_2}$, reduce to scalar quantities given by equations~\eqref{mu1_C} and~\eqref{mu2_C}, respectively.\\
\begin{figure}
\begin{center}
\includegraphics[scale=0.1]{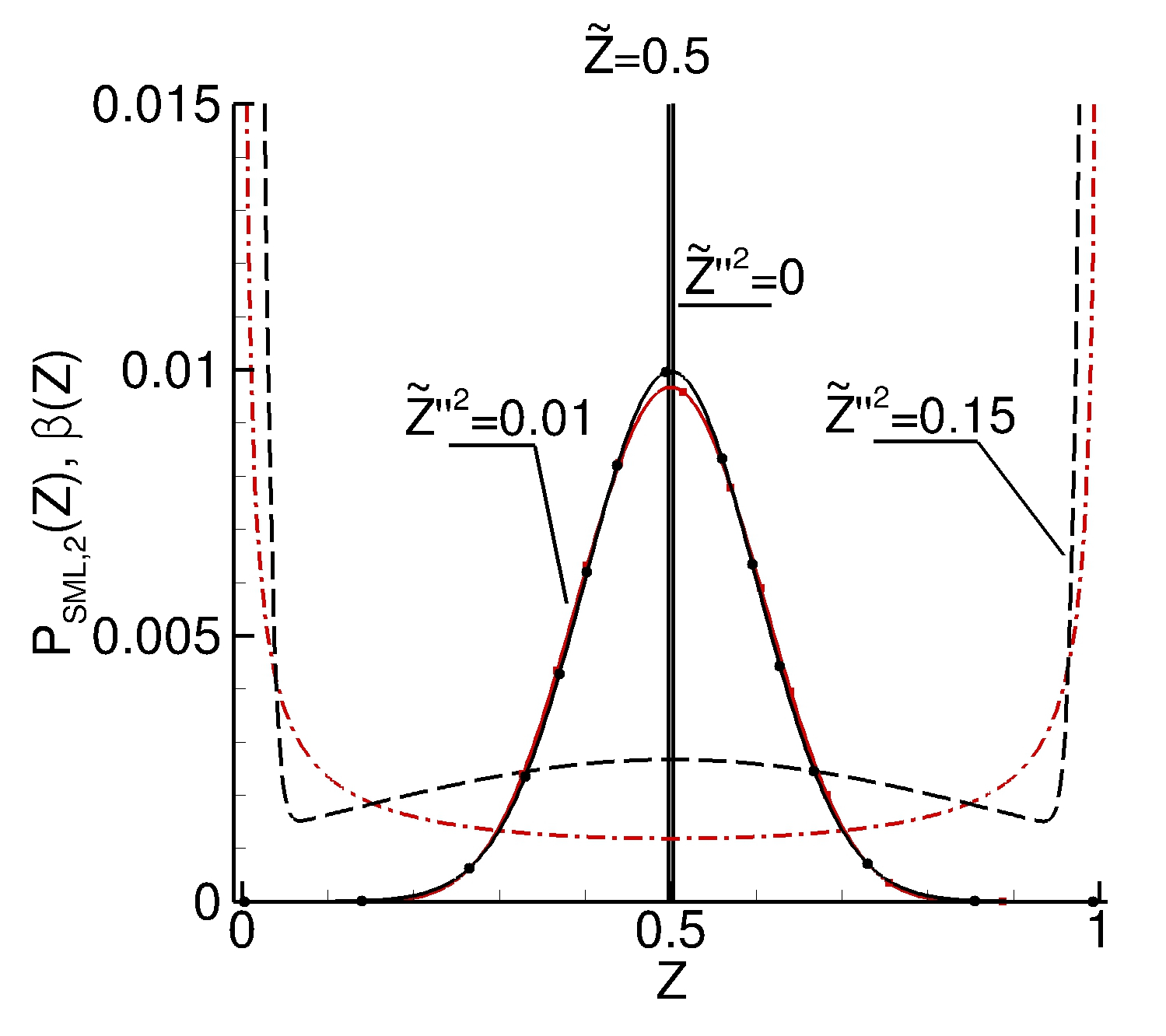}
\includegraphics[scale=0.1]{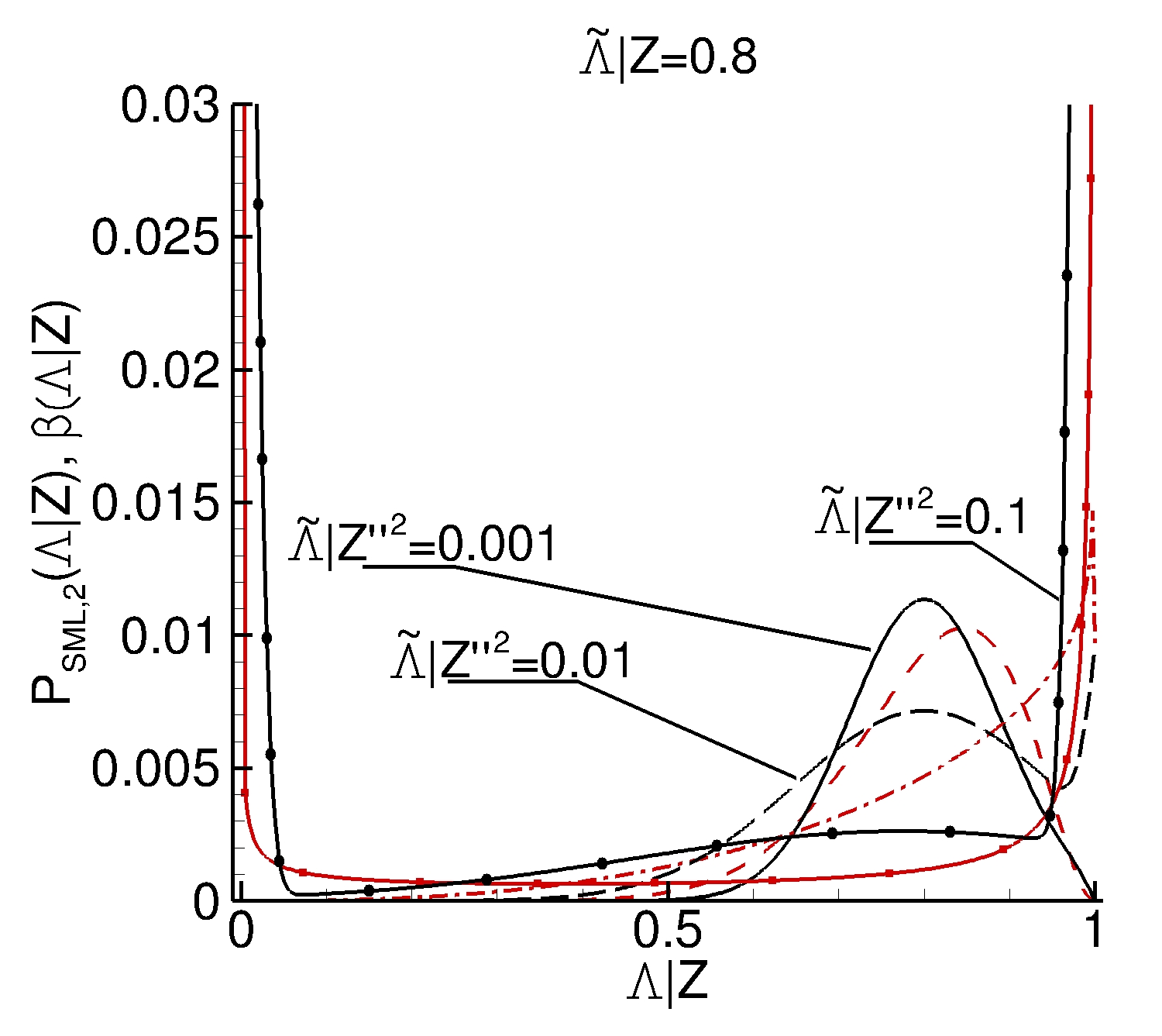}
\caption{Comparison between $\beta(Z)$ and $P_{SML,2}(Z)$, and $\beta(\Lambda|Z)$ and $P_{SML,2}(\Lambda|Z)$ distributions at two fixed arbitrary mean values, namely $\widetilde{Z}=0.5$ and $\widetilde{\Lambda|Z}=0.8$ value and for three values of the two variances, respectively.}
\label{P_var}
\end{center}
\end{figure}
Figure~\ref{P_var} shows the comparison between the two PDFs
for (arbitrarily chosen) $\widetilde{Z}=0.5$ (left panel) and $\widetilde{\Lambda|Z}=0.8$ (right panel). Three curves are plotted in each panel, corresponding to
three values of the variance in the interval $[0, \widetilde{Z}(1-\widetilde{Z})]=[0, 0.25]$ and $[0, \widetilde{\Lambda|Z}(1-\widetilde{\Lambda|Z})]=[0, 0.16]$, respectively.
One can see that the employed closure for the Lagrangian multiplier at the extrema (see equation \eqref{extrema}), does not prevent the SML distribution to  separate significantly from the $\beta$-distribution for high values of the variance. 
In fact, the two PDFs slightly differ for small and large value of the variance, the major differences being obtained for value not so close to the upper and lower limits of the admissible interval.

\section{Governing equations}
\label{goveq}

\subsection{Flow equations and numerical solution procedure}

The numerical method developed in~\cite{luigi} has been employed to solve the steady-state RANS equations with the $k$-$\omega$ turbulence closure. For multi-component reacting compressible flows, the system of governing equations can be written as:
\begin{equation}
\label{floweq}
{\partial_t \vec Q+\vec\nabla\cdot (\vec E-\vec E_{\nu})= \vec S},
\end{equation} 
where $t$ is the time. This equation is solved in an axisymmetric cylindrical reference frame, $x$ and $y$ representing the axial and the radial coordinate, respectively; 
$\vec Q$=($\overline \rho$,$\,\overline \rho \widetilde u_x$,$\,\overline \rho \widetilde u_y$,$\,\overline \rho \widetilde H$,$\,\overline \rho  k$, $\,\overline \rho \omega$,$\,\overline \rho \widetilde R_n$) is the 
vector of the conserved variables; $\overline \rho$, $(\widetilde u_x,\widetilde u_y)$, $\widetilde H$ indicate the Favre-averaged values of density, velocity components and specific total enthalpy, respectively; $k$ and $\omega$ are the turbulence kinetic energy and its specific dissipation rate; $\widetilde R_n$ is a generic set of conserved variables related to the combustion model.
In this framework, $\widetilde R_n$ is the set of independent variables of the flamelet model, namely, $\widetilde Z$, ${\widetilde{Z''^2}}$,
$\widetilde C$, ${\widetilde{C''^2}}$; $\vec E$ and $\vec E_v$ are the inviscid and viscous flux vectors \cite{dsthesis}, respectively; $\vec S$ is the vector of the source terms.\\
A cell-centred finite volume space discretization is used on a multi-block structured mesh. The convective and viscous terms are discretized by the third-order-accurate Steger and Warming~\cite{steger} flux-vector-splitting scheme and by second-order-accurate central differences, respectively. An implicit time marching procedure is used with a factorization based on the diagonalization procedure of Pulliamm and Chaussee~\cite{pulliam}, so as to allow a standard scalar alternating direction implicit (ADI) solution procedure~\cite{buelow97}. Only steady flows are dealt with in this paper. Therefore, the unsteady terms are dropped from the governing equations and the ADI scheme is iterated in the pseudo-time until a residual drop of at least five orders of magnitude for all of the conservation-law equations~\eqref{floweq} is achieved. Characteristic boundary conditions for the
flow variables are imposed at inflow and outflow points, whereas no slip and adiabatic
conditions are imposed at walls; $k$, $\omega$, and $\tilde R_n$ 
are assigned at inflow points, whilst they are linearly extrapolated at outflow points.
At solid walls, $k$ is set to zero and $\omega$ is evaluated as proposed by~\cite{menter}:
\begin{equation}
\label{eq:menter}
\omega = 60 \, \dfrac{\nu}{0.09 \, y_{n,1}^2 },
\end{equation}
where $y_{n,1}$ is the distance of the first cell center from the wall; the homogeneous Neumann
boundary condition is used for $\tilde R_n$ (non-catalytic wall).
Finally, symmetry conditions are imposed at the axis.

\subsection{Turbulent FPV transport equations}

For the case of a turbulent flame, the solution of the SLFE, namely equation~\eqref{phi}, must be written in terms of the Favre averages of $Z$ and $C$ and of their variance. 
The mean and the variance of the progress parameter are computed using equation~\eqref{normalization}.
Using model~A one can tabulate all chemical quantities in terms of $\widetilde Z$, $\widetilde{Z^{''2}}$ and $\widetilde C$ because of the properties of the $\delta$-distribution. On the other hand, models~B, C and D express $\phi$ in terms of $\widetilde{C^{''2}}$ too and therefore they need to solve a transport equation also for $\widetilde{C^{''2}}$. 
The transport equations read:
\begin{eqnarray}
\label{zmean}
 \partial_t(\overline{\rho}\widetilde{Z})+\vec\nabla\cdot(\overline{\rho}\widetilde{\vec u}\widetilde{Z})&=&
\vec\nabla\cdot\Bigl[\bigl( D+
D_{\widetilde{Z}}^t\bigr)\overline{\rho}
\vec\nabla\widetilde{Z}\Bigr],\\
\label{zvar}
\partial_t(\overline{\rho}\widetilde{Z''^2})+\vec\nabla\cdot(\overline{\rho}\widetilde{\vec u}\widetilde{Z''^2})&=&
\vec\nabla\cdot\Bigl[\bigl( D+D_{\widetilde{Z''^2}}^t\bigr)\overline{\rho}\vec\nabla\widetilde{Z''^2}\Bigr]-
\nonumber\\&-&\overline{\rho}\widetilde{\chi}+2\overline{\rho}D_{\widetilde Z}^t(\vec\nabla\widetilde{Z})^2,\\
\label{cmean}
\partial_t(\overline{\rho}\widetilde{C})+\vec\nabla\cdot(\overline{\rho}\widetilde{\vec u}\widetilde{C})&=&
\vec\nabla\cdot\Bigl[\bigl( D+D_{\widetilde{C}}^t\bigr)\overline{\rho}\vec\nabla\widetilde{C}\Bigr]+\overline{\rho}\overline{\dot\omega_C},\\
\label{cvar}
\partial_t(\overline{\rho}\widetilde{C''^2})+\vec\nabla\cdot(\overline{\rho}\widetilde{\vec u}\widetilde{C''^2})&=&
\vec\nabla\cdot\Bigl[\bigl( D+D_{\widetilde{C''^2}}^t\bigr)\overline{\rho}\vec\nabla\widetilde{C''^2}\Bigr]-
\nonumber\\&-&\overline{\rho}\widetilde{\chi_C}
+2\overline{\rho}D_{\widetilde C}^t(\vec\nabla\widetilde{C})^2+2\overline{\rho}\widetilde{C''\dot\omega''_C},
\end{eqnarray}
where $\widetilde{\chi_C}$ is modelled in terms of $\widetilde{Z''^2}$ and $\widetilde{C''^2}$~\cite{ihmeb}, namely $\widetilde{\chi_C}=\frac{\widetilde{Z''^2}\chi}{\widetilde{C''^2}}$,
$D$ is the diffusion coefficient for all of the species, given as $D=\nu/Pr$ evaluated assuming a unity Lewis number; $\nu$ and $Pr$ are the kinematic viscosity and the Prandtl number of the fluid, respectively; $D_{\widetilde Z}^t=D_{\widetilde{Z^{''2}}}^t=D_{\widetilde C}^t=D_{\widetilde{C''^2}}^t=\nu_t/Sc_{t}$ are the turbulent mass diffusion coefficients and $Sc_{t}$ the Schmidt turbulent number setted equal to $0.9$; $\dot{\omega}_{C}$ is the source term for the progress variable precomputed and tabulated in the flamelet library. The gradient transport assumption for turbulent fluxes is used and the mean scalar dissipation rate, $\widetilde{\chi}$ and $\widetilde{\chi}_C$, appear as a sink term in equations~\eqref{zvar} and \eqref{cvar}, respectively.\\
At every iteration, the values of the flamelet variables are updated using equations~\eqref{zmean}-\eqref{cvar} and the Favre-averaged thermo-chemical quantities are computed, using equation \eqref{media}. Such solutions provide the mean mass fractions which are used to evaluate all of the transport properties of the fluid and then to integrate the flow equations~\eqref{floweq}.

\section{Numerical results}
\label{results}

\begin{figure}
\begin{center}
\includegraphics[scale=0.2]{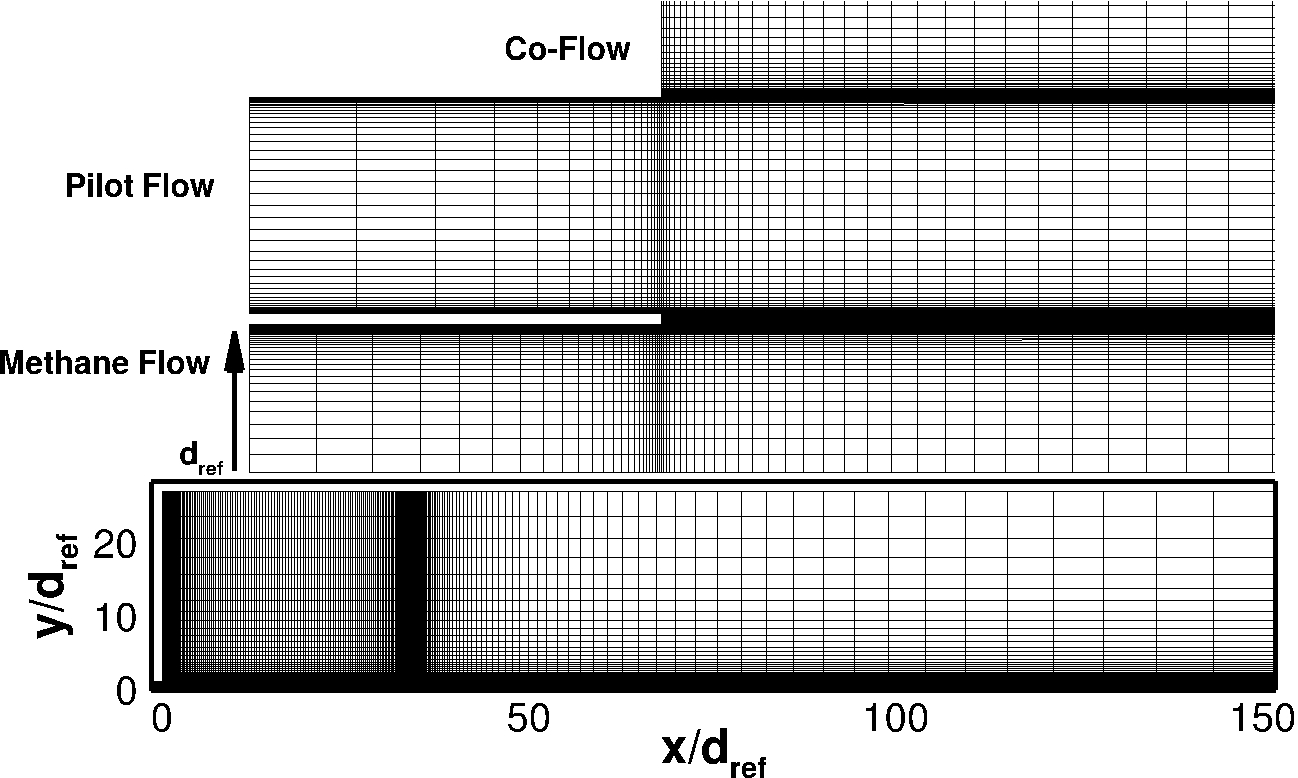}
\caption{Detail of the computational grid close to the injector (upper frame) and overall view of the computational domain (lower frame).}
\label{grid}
\end{center}
\end{figure}
This section provides the comparison among the results obtained using the four combustion models in order to assess the influence of the choice of the PDF on the prediction of turbulent non-premixed flames. The well-known Sandia flames are considered as suitable test cases, whose experimental data are available in the literature~\cite{sandia}. The steady flamelet evaluations have been performed using the FlameMaster code~\cite{flamemaster}.\\

\subsection{Sandia Flames D and E}
\begin{figure}
\begin{center}
\includegraphics[scale=0.175]{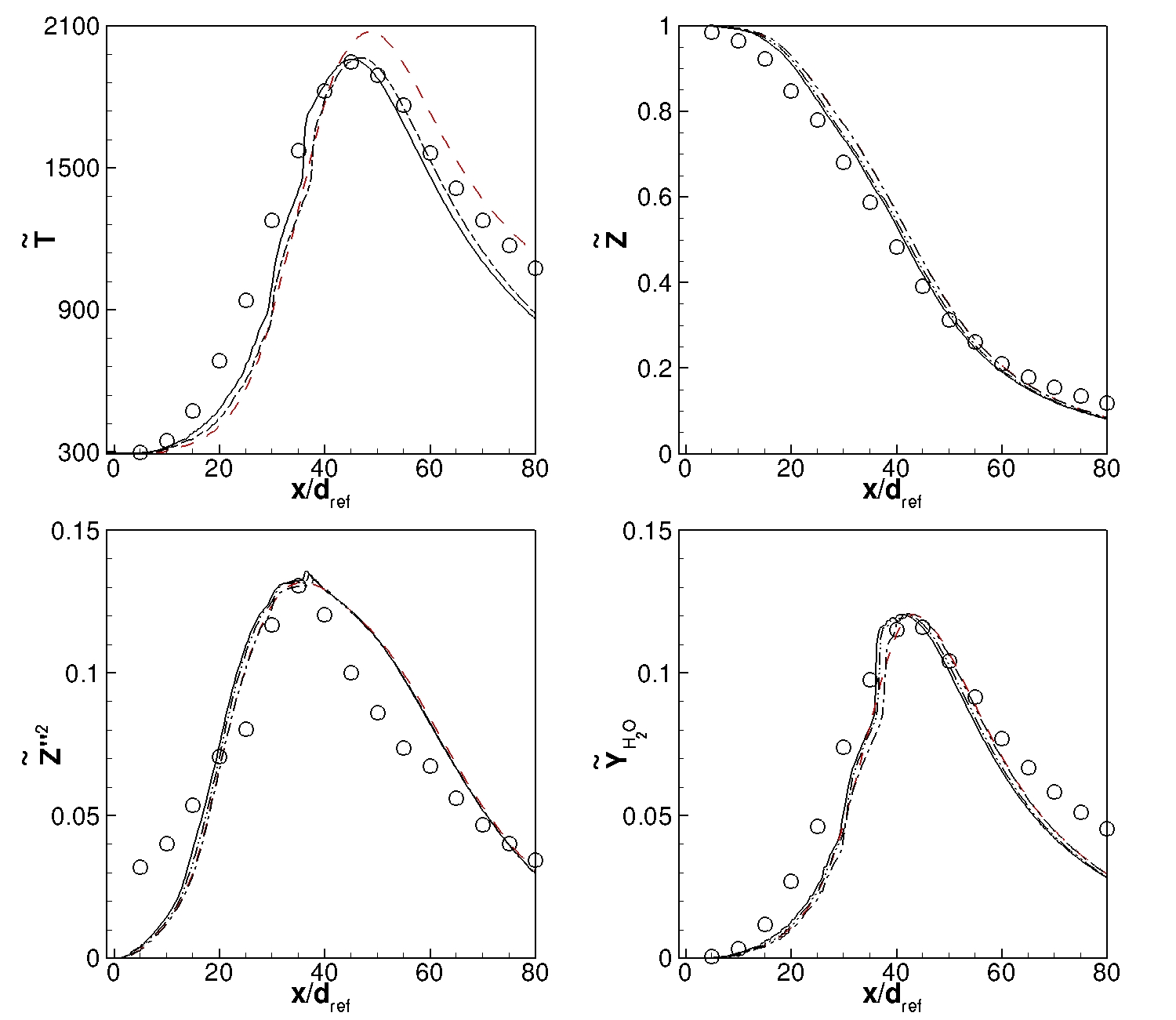}
\caption{Flame D thermo-chemical distributions along the axis $(y/d_{ref}=0)$. Model~A, red dashed line; Model~B, dashed-dotted line; Model~C, dashed-dotted-dotted line; Model~D, solid line; Symbols, experimental data.}
\label{D_cent}
\end{center}
\end{figure}
\begin{figure}
\begin{center}
\includegraphics[scale=0.175]{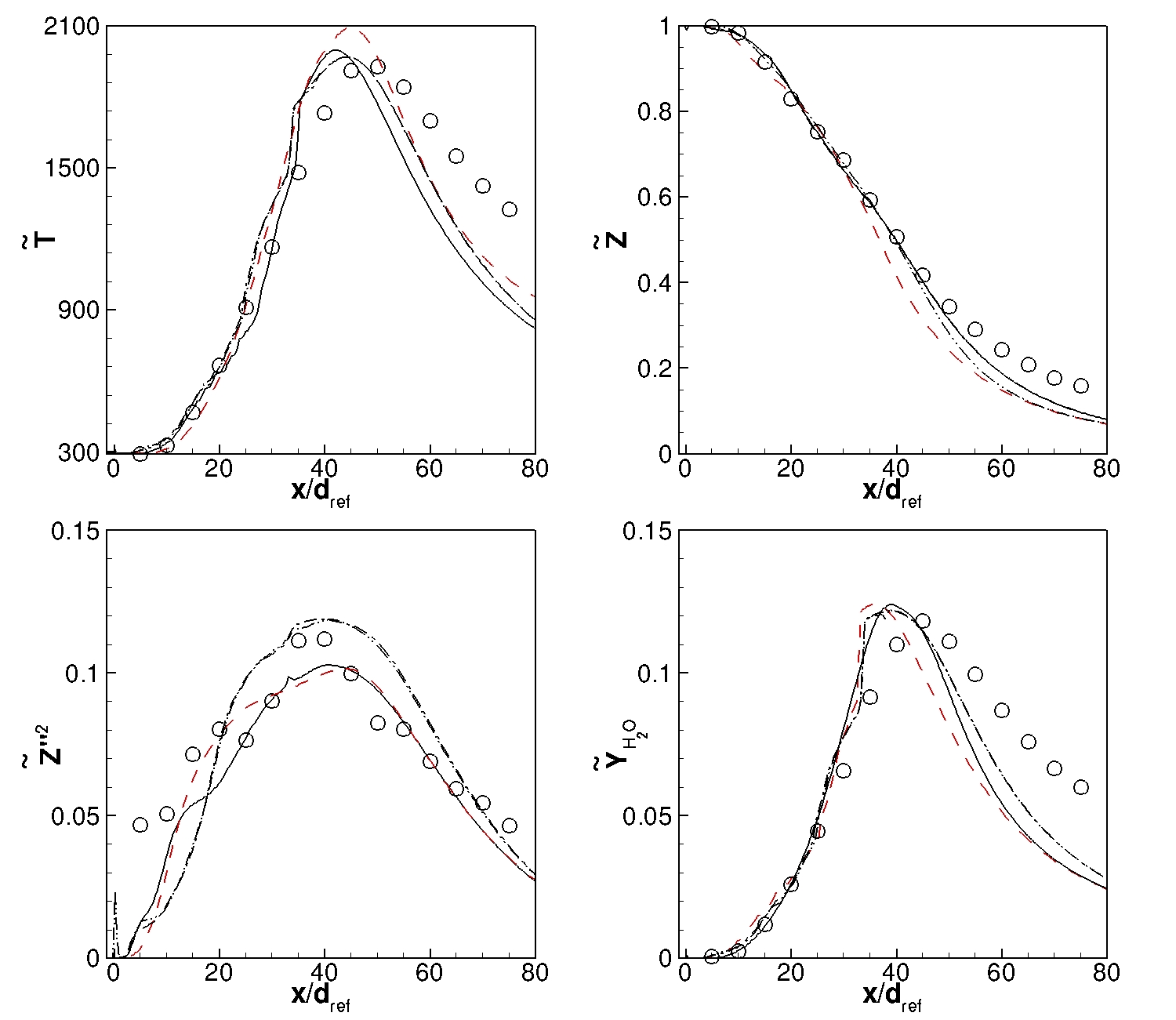}
\caption{Flame E thermo-chemical distributions along the axis $(y/d_{ref}=0)$. Model~A, red dashed line; Model~B, dashed-dotted line; Model~C, dashed-dotted-dotted line; Model~D, solid line; Symbols, experimental data.}
\label{E_cent}
\end{center}
\end{figure}
In the Sandia flame experiments, piloted partially premixed CH$_4$/air diffusion flames at the same pressure, equal to $100.6\ kPa$, and 
at different Reynolds numbers have been studied by Barlow \& Frank~\cite{barlow98,sandia}. The Reynolds number, $Re$, is based on the nozzle diameter, the jet bulk velocity, and the kinematic viscosity of the fuel. The diameter of the nozzle of the central jet is $d_{ref}=7.2$~mm and the inner and outer diameters of the annular pilot nozzle are equal to  $7.7$~mm and $18.2$~mm, respectively. 
The oxidizer air ($Y_{O_2}=0.233$, $Y_{N_2}=0.767$) is supplied as a co-flow at 291 K.
The main fuel jet consists of a mixture of methane and air with a volumetric
ratio equal to 1:3, and a stoichiometric mixture fraction $Z_{st}=0.351$; the corresponding
mass fractions are $Y_{CH_4}=0.156$, $Y_{O_2}=0.196$, and $Y_{N_2}=0.648$.
The pilot stream is a lean premixed gas mixture with an equivalence ratio of $\phi=0.77$, corresponding to the equilibrium
composition with $Y_{CO_2}=0.110$, $Y_{O_2}=0.056$, $Y_{N_2}=0.734$, $Y_{OH}=0.002$, and $Y_{H_2O}=0.098$.
The inlet value of the progress variable is equal to $0.208$. Flame D ($Re=22400$) presents very low degree of local extinction, whereas Flame E ($Re =33600$) has significant and increasing probability of local extinction near the pilot.
\begin{table}[!tb]
\centering
\caption{Conditions for Sandia D-E flame experiment.}
\label{tab:sandiaD}
\begin{tabular}{l|c c c c c c}
\hline
& $u$ (m/s) & $T$ (K)& $\widetilde Z$& $\widetilde{Z''^2}$& Tu$_{in}$& $\lambda_{T,in}$ (mm)\\
Main jet &49.6 $\pm$ 2 (D)&294.0&1&0&5\%&0.5\\
&74.4 $\pm$ 2 (E)&&&&&\\
\\
Pilot stream& 11.4 $\pm$ 0.5 (D)&1800&0.27&0.0075&5\%&0.5\\
&17.1 $\pm$ 0.75 (E)&&&&&\\
\\
Co-flow&0.9 $\pm$ 0.05&291.0&0&0&0.5\%&1.0\\ 
\hline
\end{tabular}
\end{table}
The experimental values of the axial velocities and temperatures have been assigned at the inlet points (left boundary) together
with the turbulent intensity levels, Tu$_{in}$, and turbulent characteristic length scale, $\lambda_{T,in}$,
as reported in Table~\ref{tab:sandiaD}.
Free-slip conditions have been imposed at the points on the upper boundary.
The computational domain, shown in figure~\ref{grid}, is axisymmetric and includes a part of the burner; it has a length of $150\ d_{ref}$ and $27\ d_{ref}$ along the axial and radial directions, respectively, and has been discretized using about $47000$ cells. 
Computations have been performed using the combustion scheme described by the GRIMECH~$3.0$~\cite{grimech30}: $325$ sub-reactions upon $53$ species. The flamelet library is computed over a grid with $125$ uniformly distributed points in the $\widetilde Z$ and $\widetilde C$ directions and $25$ uniformly distributed points in the $\widetilde{Z''^2}$ and $\widetilde{C''^2}$ directions.

Experimental data for the time-averaged major species and temperature are available 
as radial distributions at several stream-wise 
locations and as axial distributions along the center-line~\cite{sandia}. 
The statistical errors of the measurements are below $10\%$ for the CO mass fraction, and below $5\%$ for the other quantities.
\begin{figure}
\begin{center}
\includegraphics[scale=0.11]{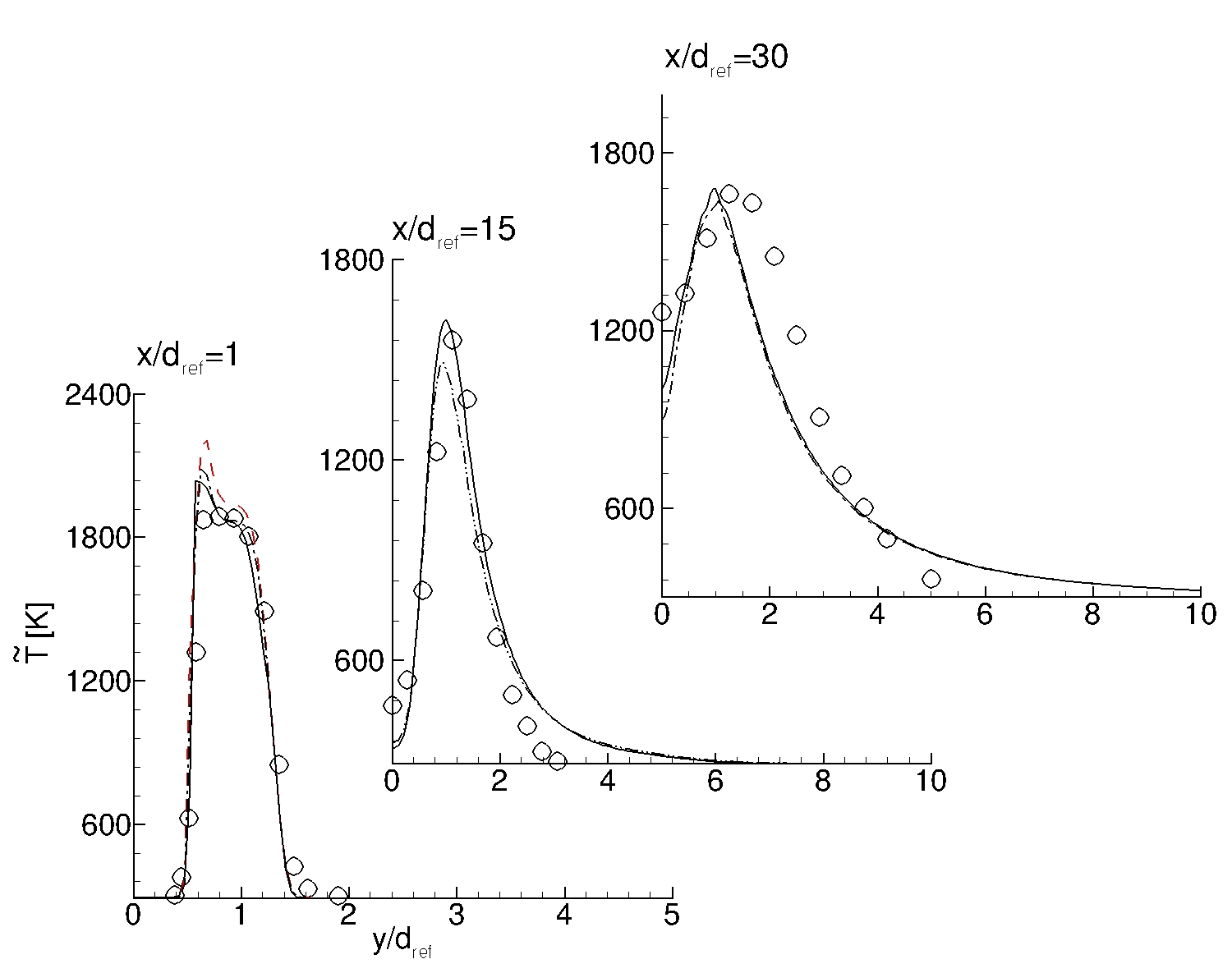}
\includegraphics[scale=0.11]{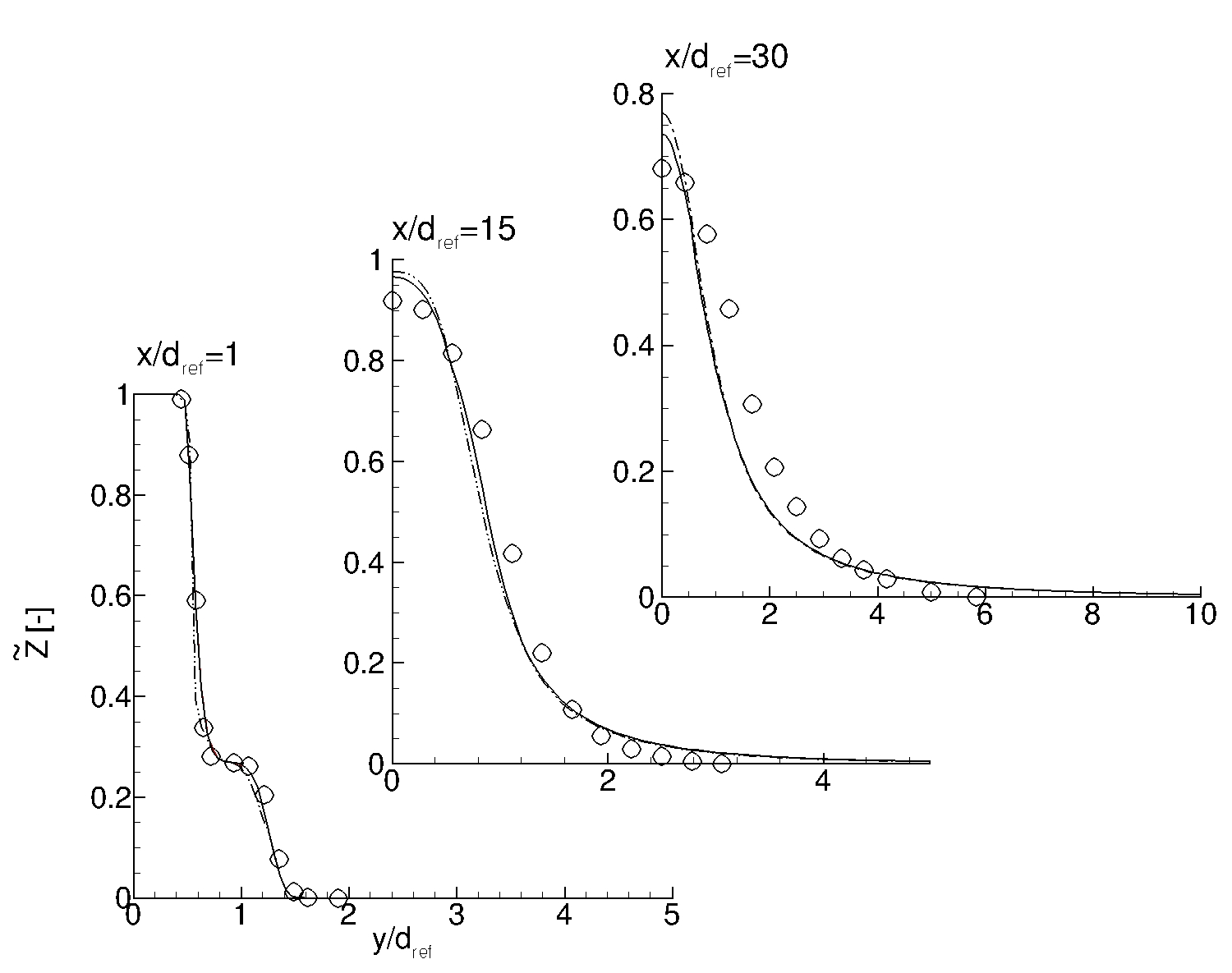}\\
\includegraphics[scale=0.11]{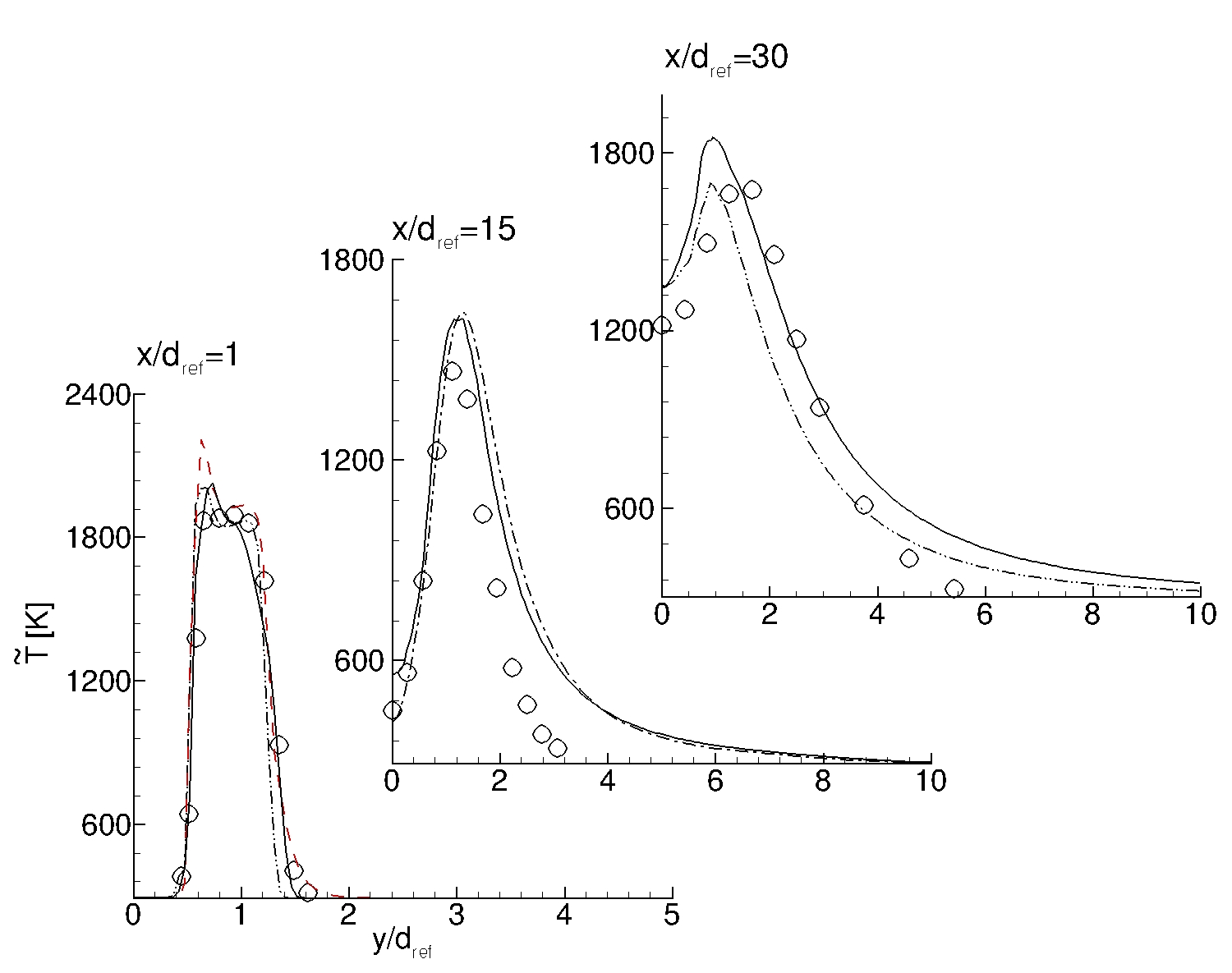}
\includegraphics[scale=0.11]{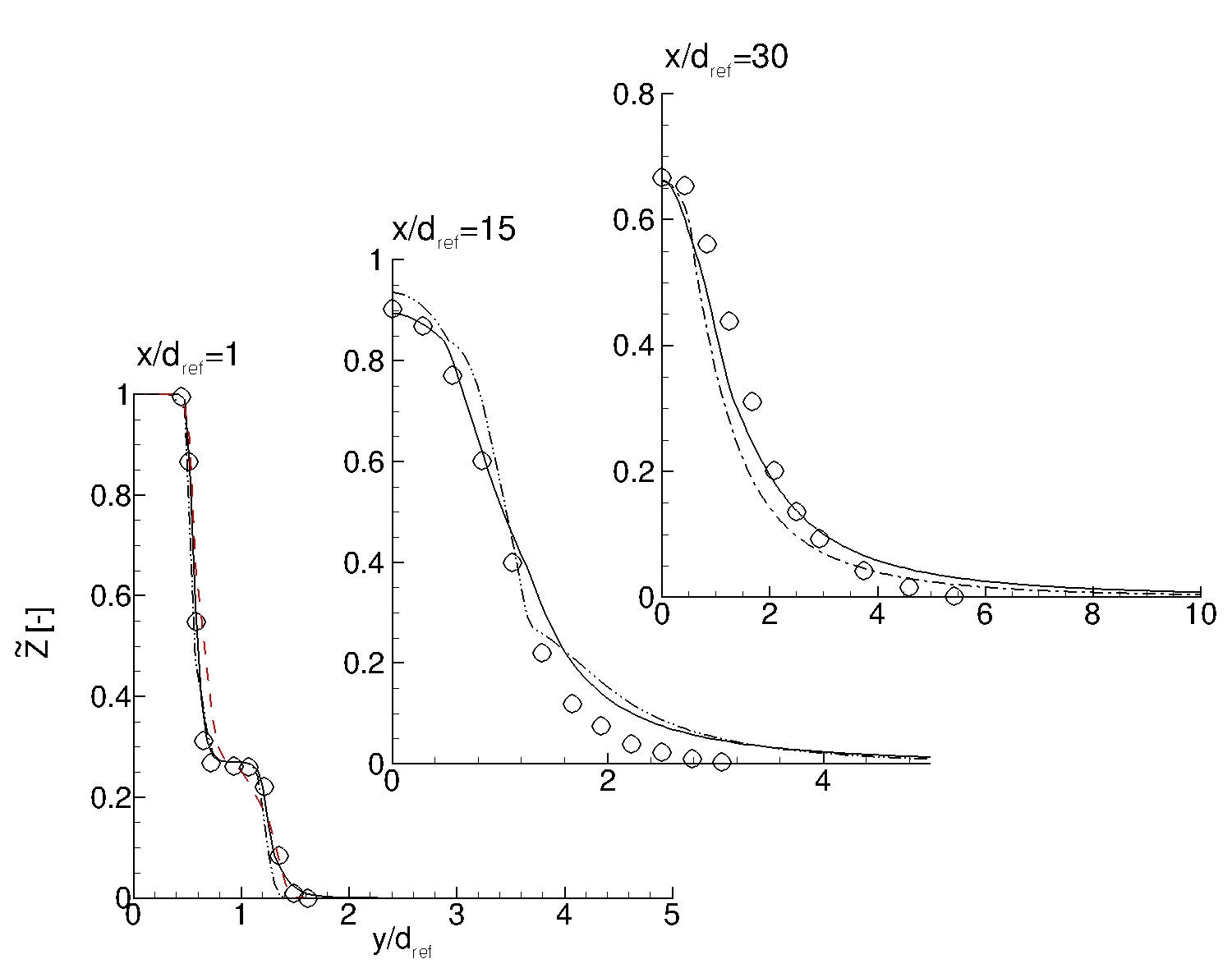}\\
\caption{Flame~D (upper frame) and~E (lower frame) thermo-chemical distributions taken at, $x/d_{ref}=1$, $x/d_{ref}=15$, and, $x/d_{ref}=30$ from top to bottom. Model~A, shown only in the first section corresponds to the red-dashed line; Model~D, black-solid line; Symbols, experimental data.}
\label{rad}
\end{center}
\end{figure}
%
%%%%%%%\begin{figure}
%%%%%%%\begin{center}
%%%%%%%\includegraphics[scale=0.15,angle=-90]{E_radial.jpg}
%%%%%%%\caption{Flame E thermo-chemical distributions taken at, $x/d_{ref}=1$, $x/d_{ref}=15$, $x/d_{ref}=30$, and $x/d_{ref}=45$ from top to bottom. Model~A, red-dashed line; Model~D, black-solid line; Symbols, experimental data.}
%%%%%%%\label{E_rad}
%%%%%%%\end{center}
%%%%%%%\end{figure}
%
Figures~\ref{D_cent} and~\ref{E_cent} show the comparison among the distributions of $\widetilde T$, $\widetilde{Z}$, $\widetilde{Z''^2}$, and $\widetilde{Y}_{H_2O}$ obtained with the four FPV models along the flame center-line for flame D and flame E, respectively. 
All models provide temperature distributions in reasonable agreement with the experimental data, model~A predicting a peak value of the temperature remarkably higher than the others.
Moreover, for flame~E a significant shift of the location of temperature peak is obtained with all models, probably due to the higher inlet velocity with respect to the flame~D. 
The results of model~B and model~C are almost always overlapped. On the other hand, model~D shows a slight improvement with respect to the others. 
This two figures show that all of the models are able to provide a good estimation of the mixture fraction distribution. Moreover the evaluation of the $H_2O$ mass fraction is in good agreement with the experimental data and here the four models present very small differences since they are based on the same efficient kinetic mechanism~\cite{grimech30}. 
The distributions of temperature and mixture fraction obtained with model~C and model~D at three axial sections are shown in figure~\ref{rad} for flame~D and~E.  
The comparison with model~A is provided only for the first section, where a large discrepancy is observed in the temperature profile. 
In fact, one can see that model~A predicts a maximum temperature of about $2200\ K$ at $x/d_{ref}=1$ whereas the maximum experimental value is about $1900\ K$; in contrast, models~C and~D provide an improved predicted value of about $2000\ K$.
Models~C and~D provide results in acceptable agreement with the experimental data, model~D being slightly more accurate.   
Concerning the distribution of $\widetilde{Z}$, an overall good agreement is achieved between the experimental data and the numerical results.
The results obtained are in agreement with the results obtained using several combustion models using either RANS or LES  approaches~\cite{ochoa2012,zoller2011,kemenov2011,ferraris2008}.
\subsection{Mixture fraction-conditioned results}
\begin{figure}
\begin{center}
\includegraphics[scale=0.2]{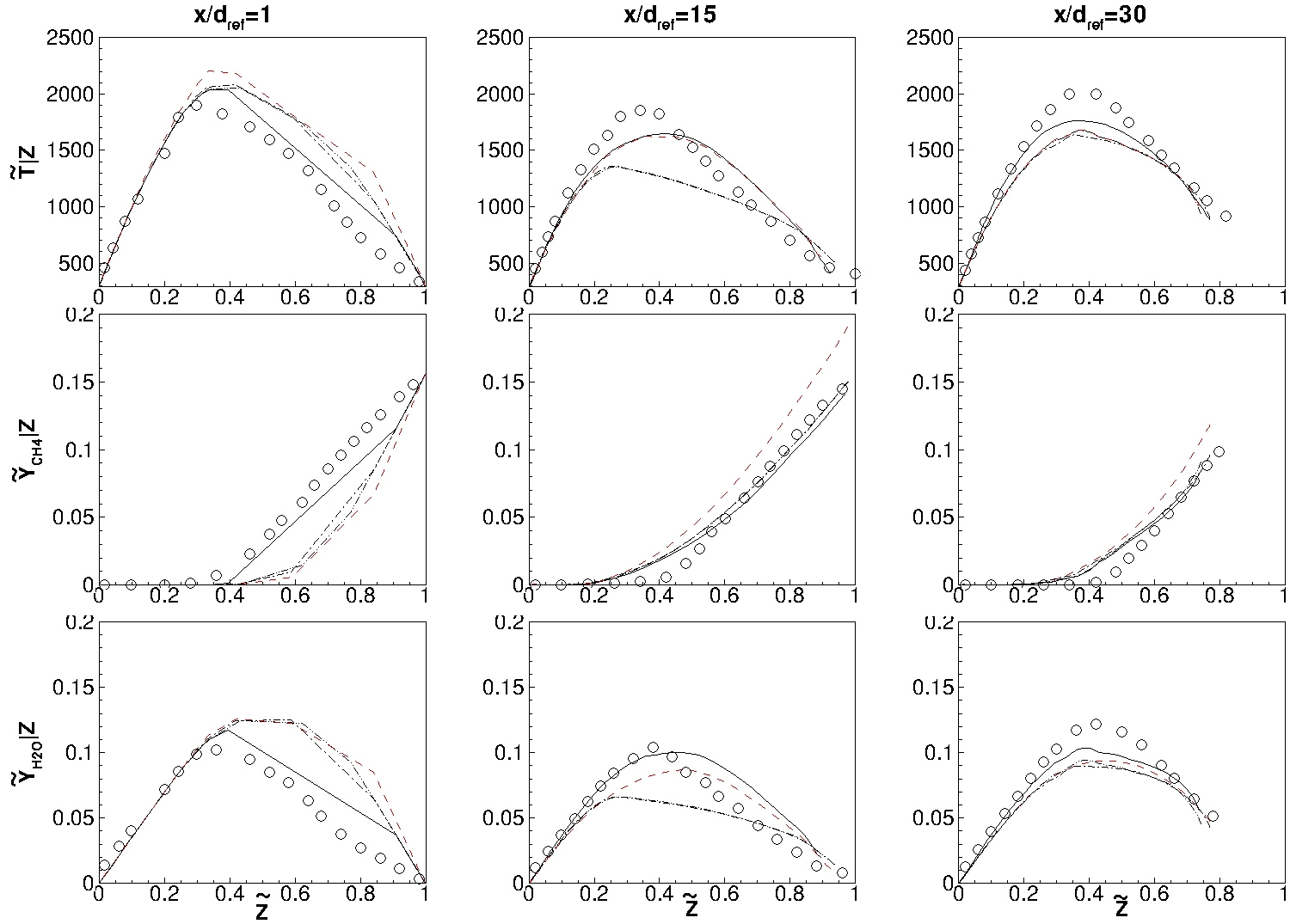}
\caption{Flame D thermo-chemical $Z$-conditioned distributions taken at, $x/d_{ref}=1$, $x/d_{ref}=15$, and $x/d_{ref}=30$ respectively. Model~A, red-dashed line; Model~B, dashed-dotted line; Model~C, dashed-dotted-dotted line; Model~D, black-solid line; Symbols, experimental data.}
\label{D_rad_cond}
\end{center}
\end{figure}
\begin{figure}
\begin{center}
\includegraphics[scale=0.2]{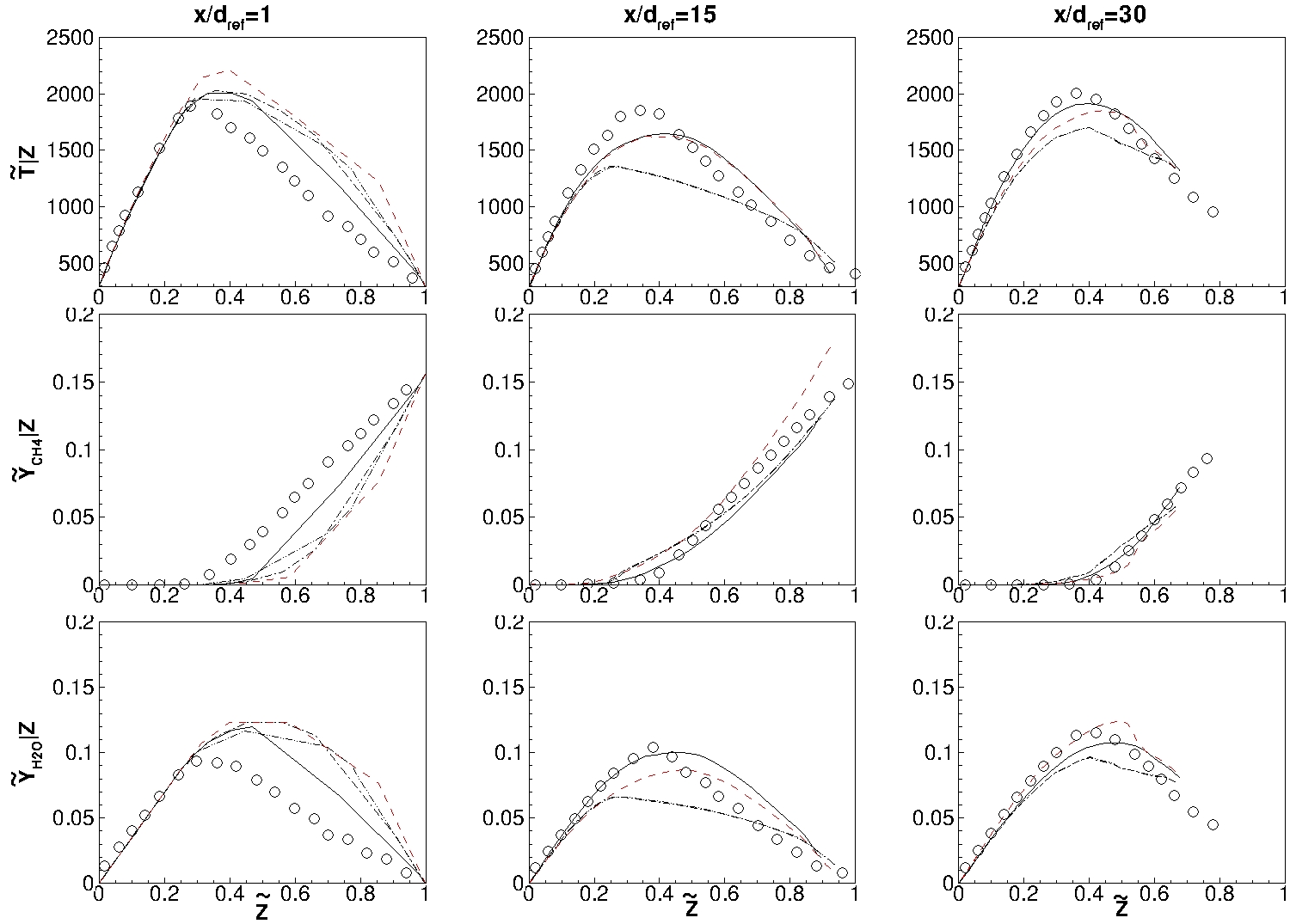}
\caption{Flame E thermo-chemical $Z$-conditioned distributions taken at, $x/d_{ref}=1$, $x/d_{ref}=15$, and $x/d_{ref}=30$ respectively. Model~A, red-dashed line; Model~B, dashed-dotted line; Model~C, dashed-dotted-dotted line; Model~D, black-solid line; Symbols, experimental data.}
\label{E_rad_cond}
\end{center}
\end{figure}
For an assessment of the combustion-model performance, a flow-field-independent comparison with the experimental data~\cite{sandia} would be needed. 
This can be obtained approximatively by
analysing the Favre-averaged mixture fraction-conditioned thermo-chemical quantities obtained for both flames.
Figure~\ref{D_rad_cond} shows the conditioned temperature, methane, and water mass fractions distributions along three radial sections taken at $d_{ref}$, $15 d_{ref}$ and $30 d_{ref}$, for flame D. 
Model~D provides results in very good agreement with the experimental data~\cite{sandia} and always in better agreement with respect to the other three models.
Focusing on the first section, one can see the remarkable differences between model~A and model~D; model~A overestimates the methane consumption and, consequently, the temperature and the water productions. 
The differences fade through the three sections: the maximum temperature difference is about $395 K$ at $\widetilde Z=0.85$ in the first section, whereas, it is about $137 K$ at $\widetilde Z=0.215$ in the third section.
The same conclusions are obtained analysing figure~\ref{E_rad_cond} showing the $Z$-conditioned quantities at the same three sections for the flame~E. 
It is noteworthy that all of the models overestimate the fuel consumption in the fuel rich side, even if the results are more accurate for model~D. 
Such a behaviour is probably due to the model construction neglecting the interaction between turbulent mixing, chemistry, and radiative heat transfer~\cite{ihmeb}. 
\begin{figure}
\begin{center}
\includegraphics[scale=0.2]{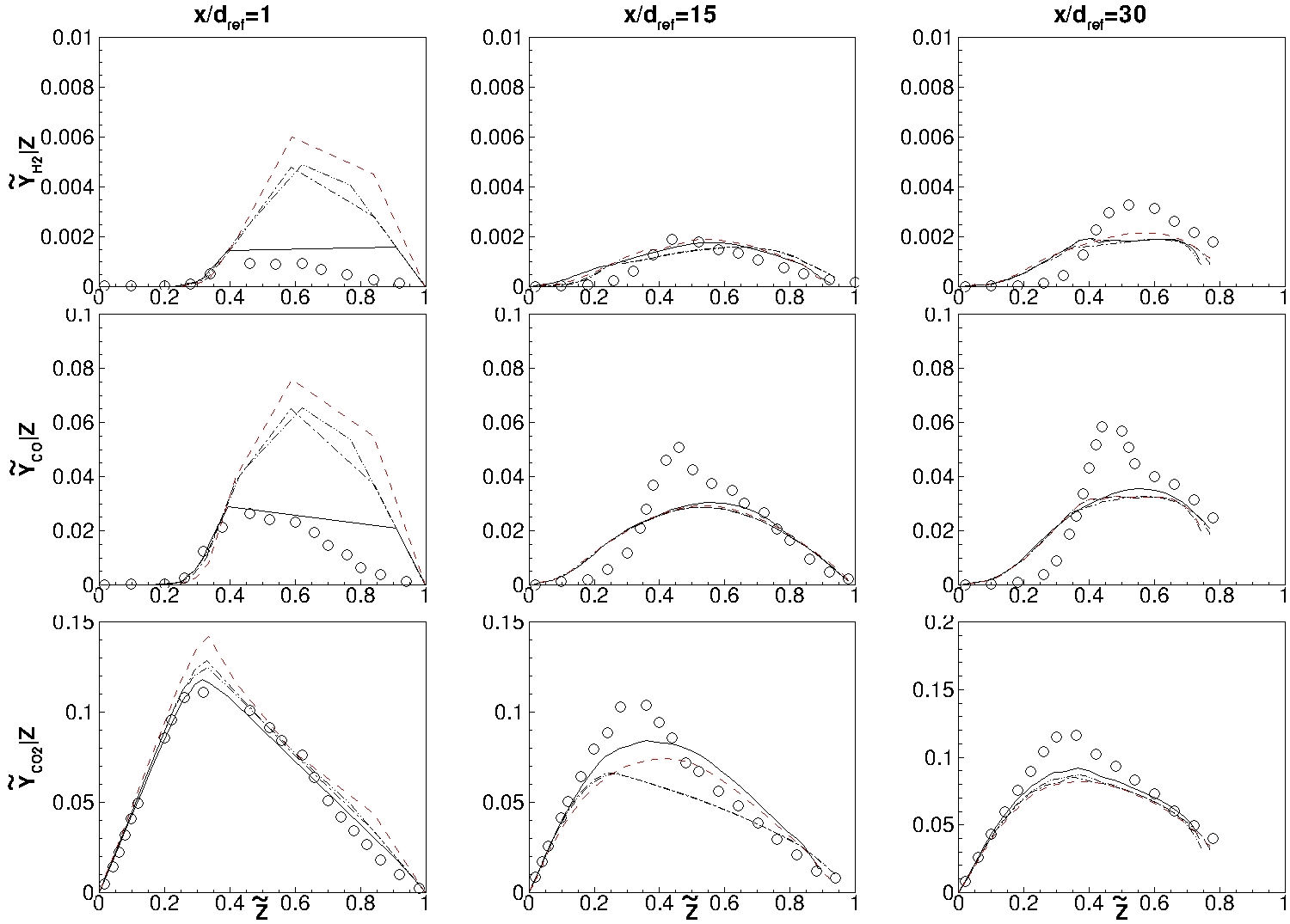}
\caption{Flame D chemical species mass fractions $Z$-conditioned distributions taken at, $x/d_{ref}=1$, $x/d_{ref}=15$, and $x/d_{ref}=30$ respectively. Model~A, red-dashed line; Model~B, dashed-dotted line; Model~C, dashed-dotted-dotted line; Model~D, black-solid line; Symbols, experimental data.}
\label{D_rad_cond_chem}
\end{center}
\end{figure}
\begin{figure}
\begin{center}
\includegraphics[scale=0.2]{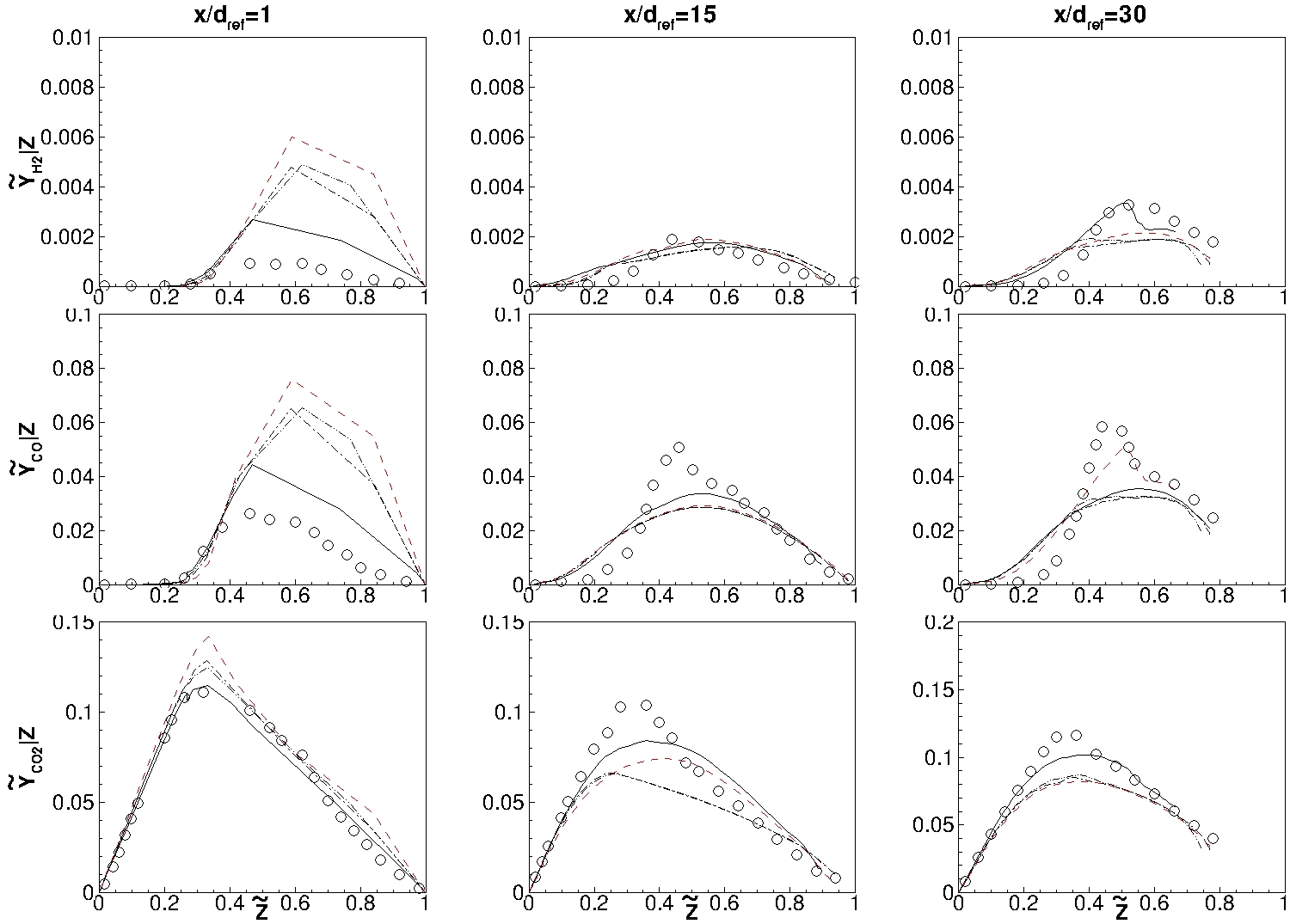}
\caption{Flame E chemical species mass fractions $Z$-conditioned distributions taken at, $x/d_{ref}=1$, $x/d_{ref}=15$, and $x/d_{ref}=30$ respectively. Model~A, red-dashed line; Model~B, dashed-dotted line; Model~C, dashed-dotted-dotted line; Model~D, black-solid line; Symbols, experimental data.}
\label{E_rad_cond_chem}
\end{center}
\end{figure}
Figures~\ref{D_rad_cond_chem} and \ref{E_rad_cond_chem} show the $H_2$, $CO$ and $CO_2$ $Z$-conditioned mass fractions, for flames D and E, respectively. 
Remember that the sum of the mass fractions of the three species together with the water mass fraction define the progress variable (see equation~\eqref{progress_variable}). 
The figures are provided to check the model consistency with the physical phenomenon observed; in fact,
one can see that in the farthest sections, where the reactions are fading (the progress variable tends to zero),
the differences between the models are small with respect to the differences observed in the sections close to the burner.
Finally, this analysis shows that all of th models provide quite good capability in the lean-fuel region of the flame, $\widetilde Z\leq 0.3$.
In the fuel-rich side, model~D gives improved results, while slightly overestimate the temperature and the water productions 
due to the over prediction of the methane consumption, whereas model~A provides unsatisfactory results. In this context it is remarked that even if model~D is not able to replicate the accuracy of transported-PDF models, see e.g.~\cite{juddoo2011}, its results appear to provide a remarkable improvement with respect to other FPV models.

\subsection{About the joint distribution of $Z$ and $\Lambda$}
\begin{figure}
\begin{center}
\includegraphics[scale=0.055]{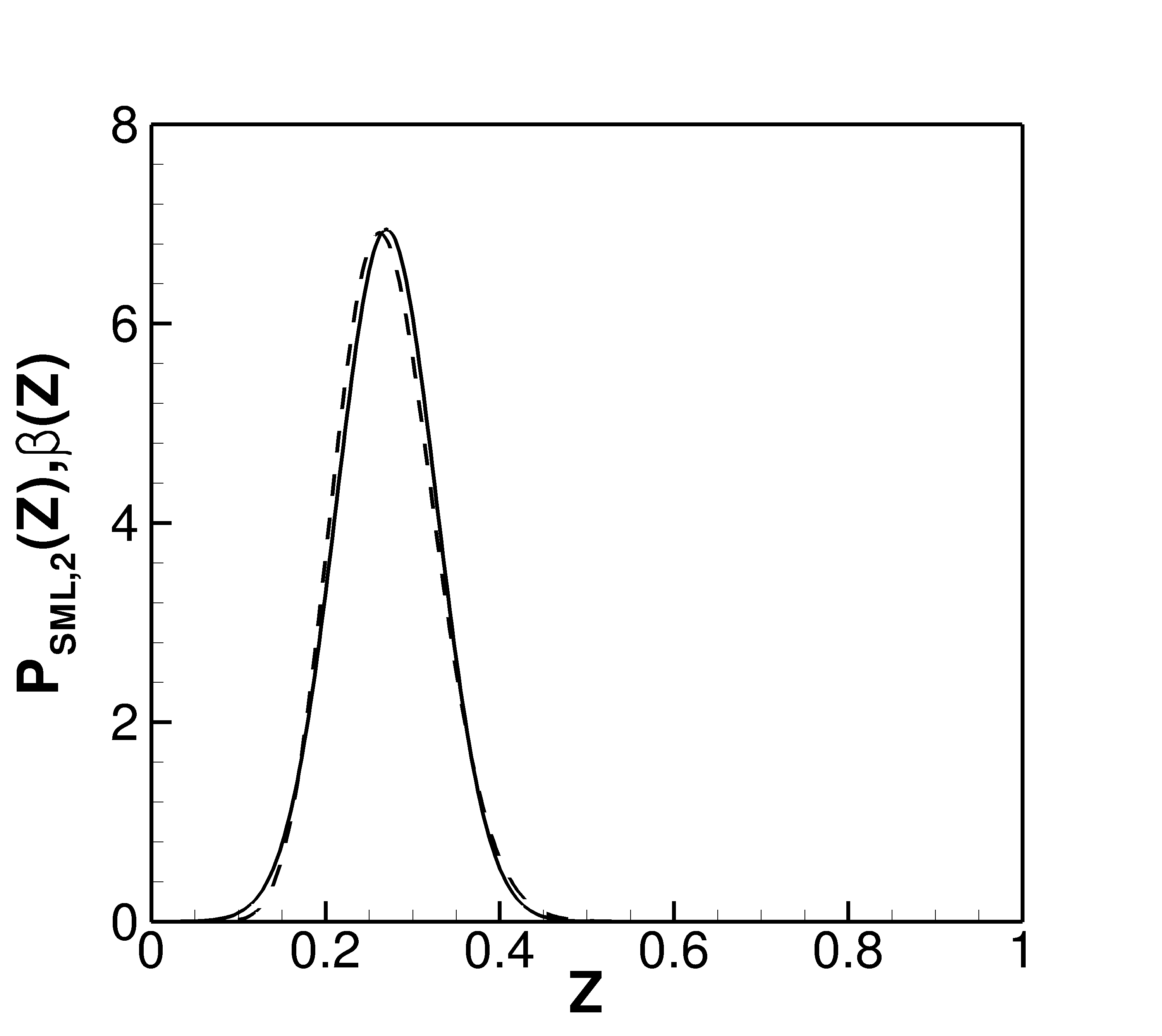}
\includegraphics[scale=0.055]{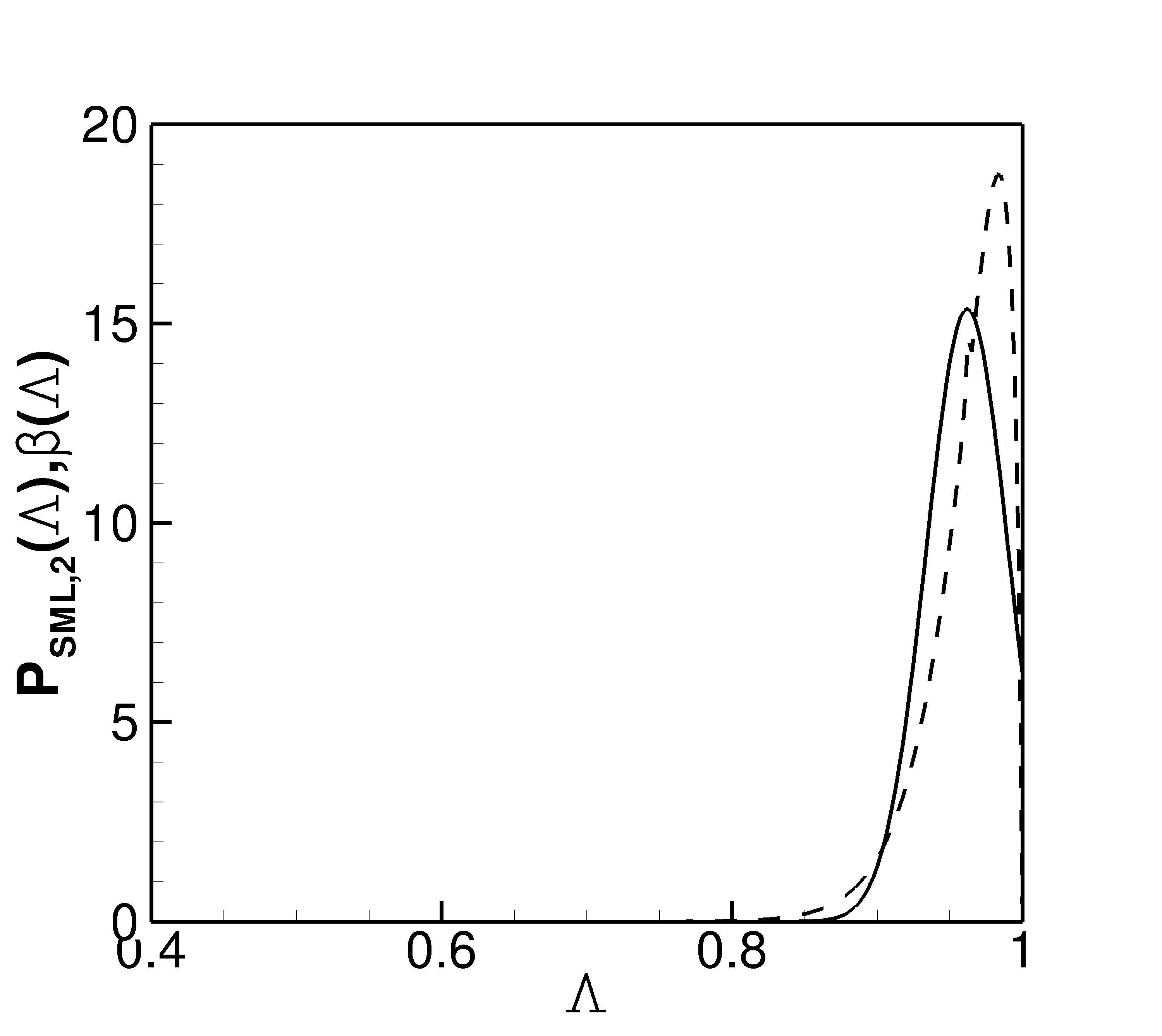}
\caption{Distribution of $Z$ and $\Lambda$ at the point $(x/d_{ref},y/d_{ref})=(0,1)$. The solid line is the $P_{SML,2}$ distribution and the dashed one the $\beta$-distribution.}
\label{x_0_y_1}
\end{center}
\end{figure}
\begin{figure}
\begin{center}
\subfigure[$(x/d_{ref},y/d_{ref})=(0,1)$]{\includegraphics[scale=0.055]{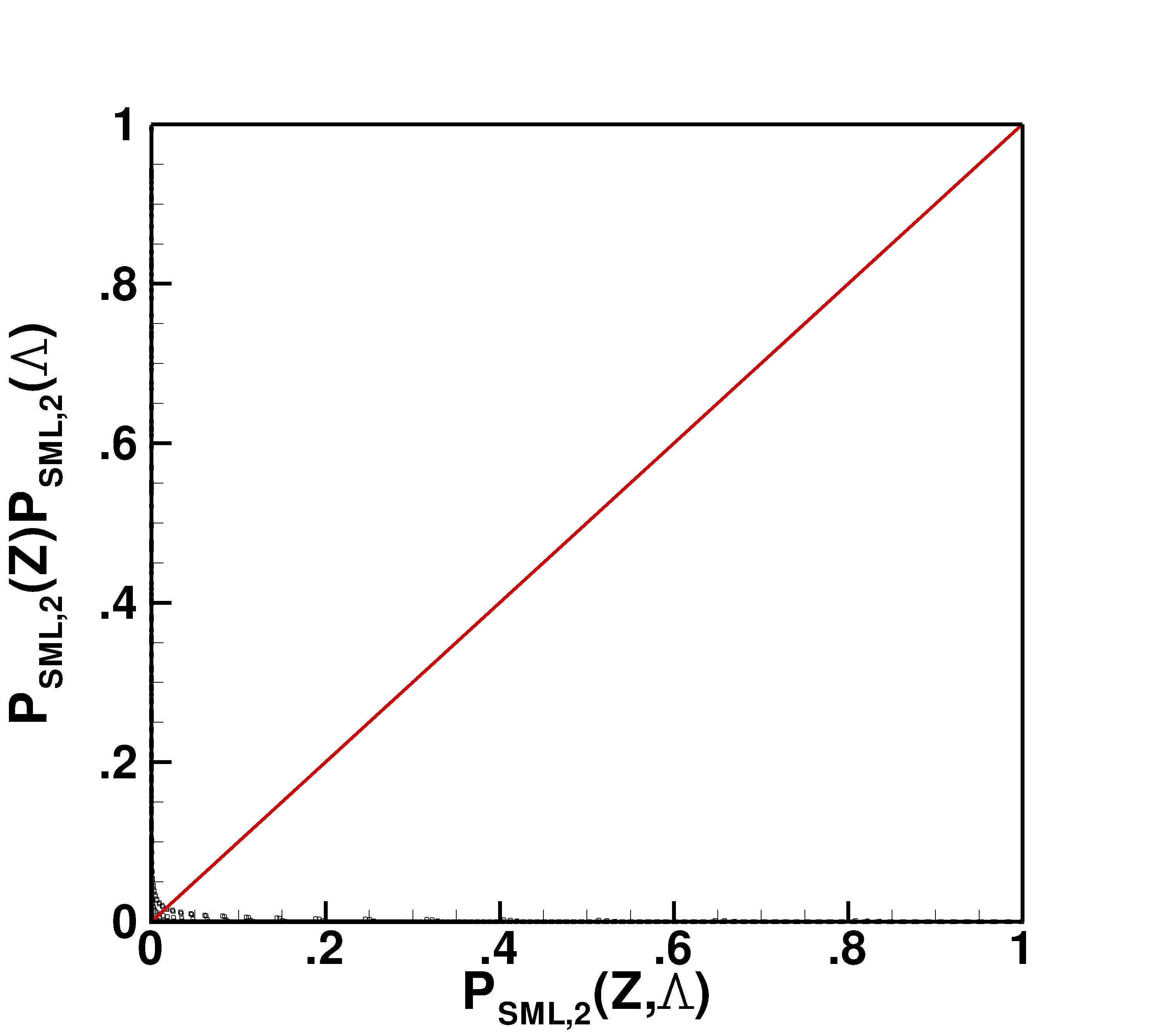}}
\subfigure[$(x/d_{ref},y/d_{ref})=(20.85,1)$]{\includegraphics[scale=0.11]{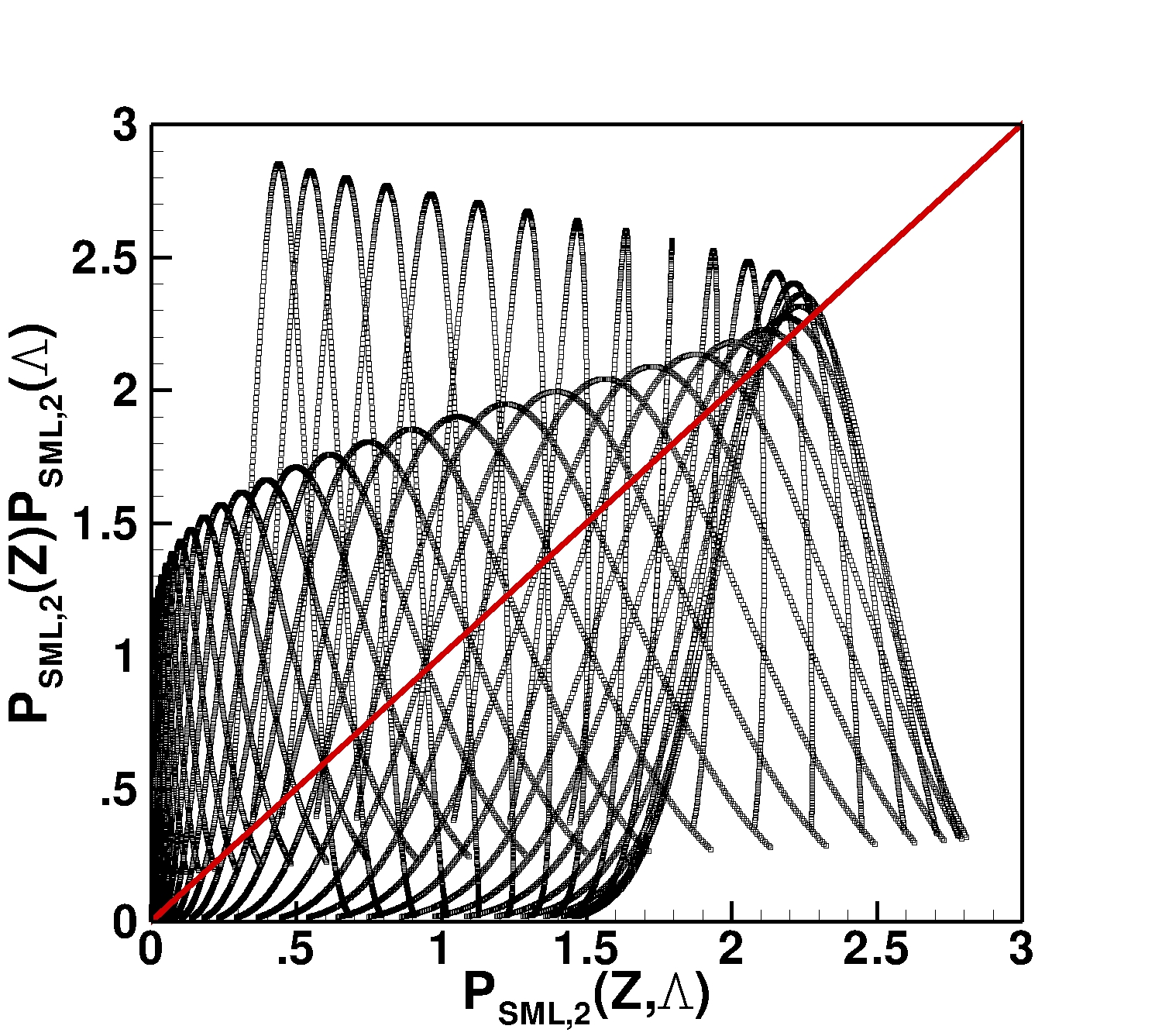}}
\caption{Scatter-plot of the joint PDF with statistical independence hypothesis,$P_{SML,2}(Z)P_{SML,2}(\Lambda)$, versus the joint PDF, $P_{SML,2}(Z,\Lambda)$, at two different point in the Flame~E flow field. The solid line is the bisector.}
\label{scat}
\end{center}
\end{figure}
\begin{figure}
\begin{center}
\includegraphics[scale=0.055]{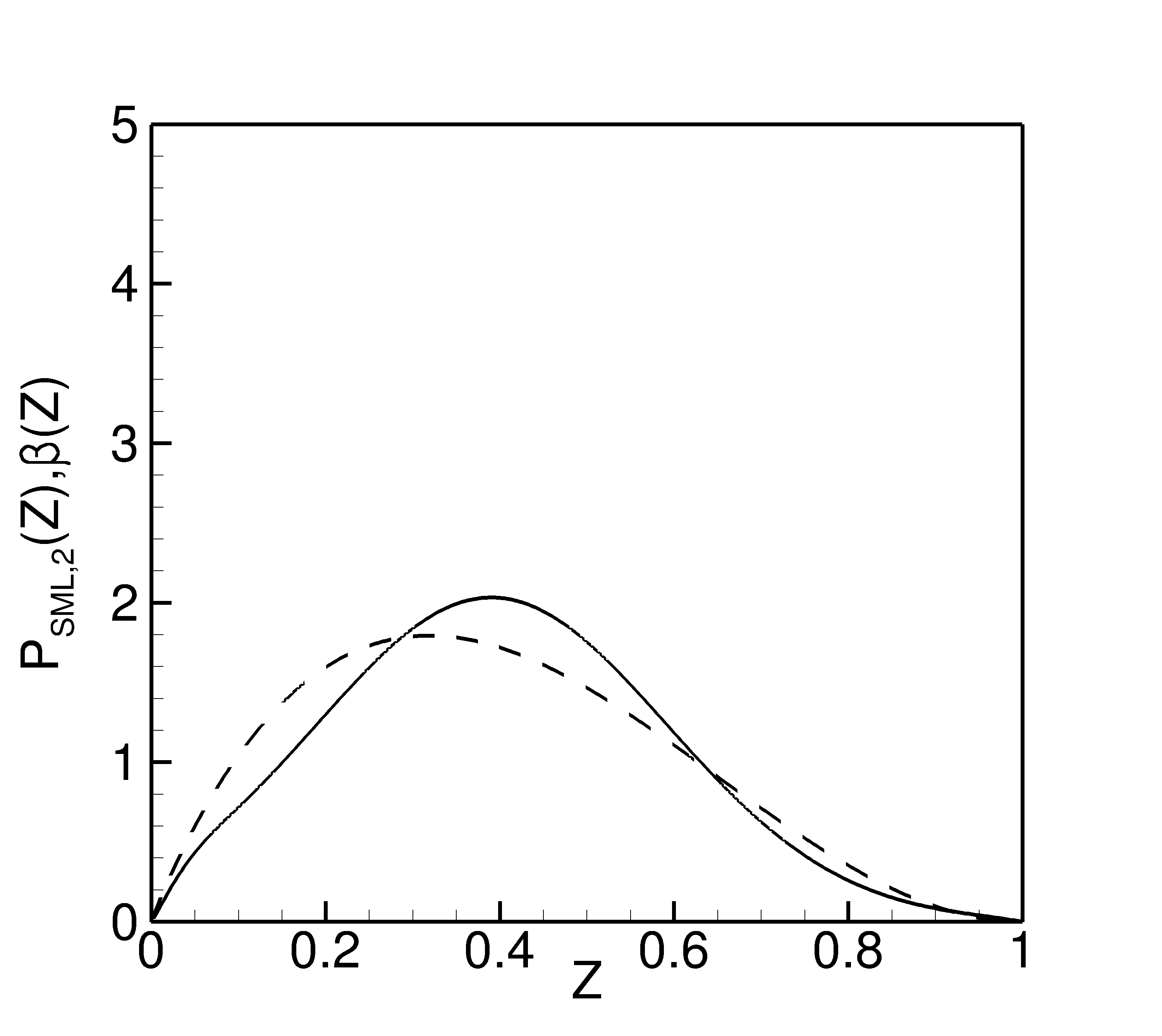}
\includegraphics[scale=0.055]{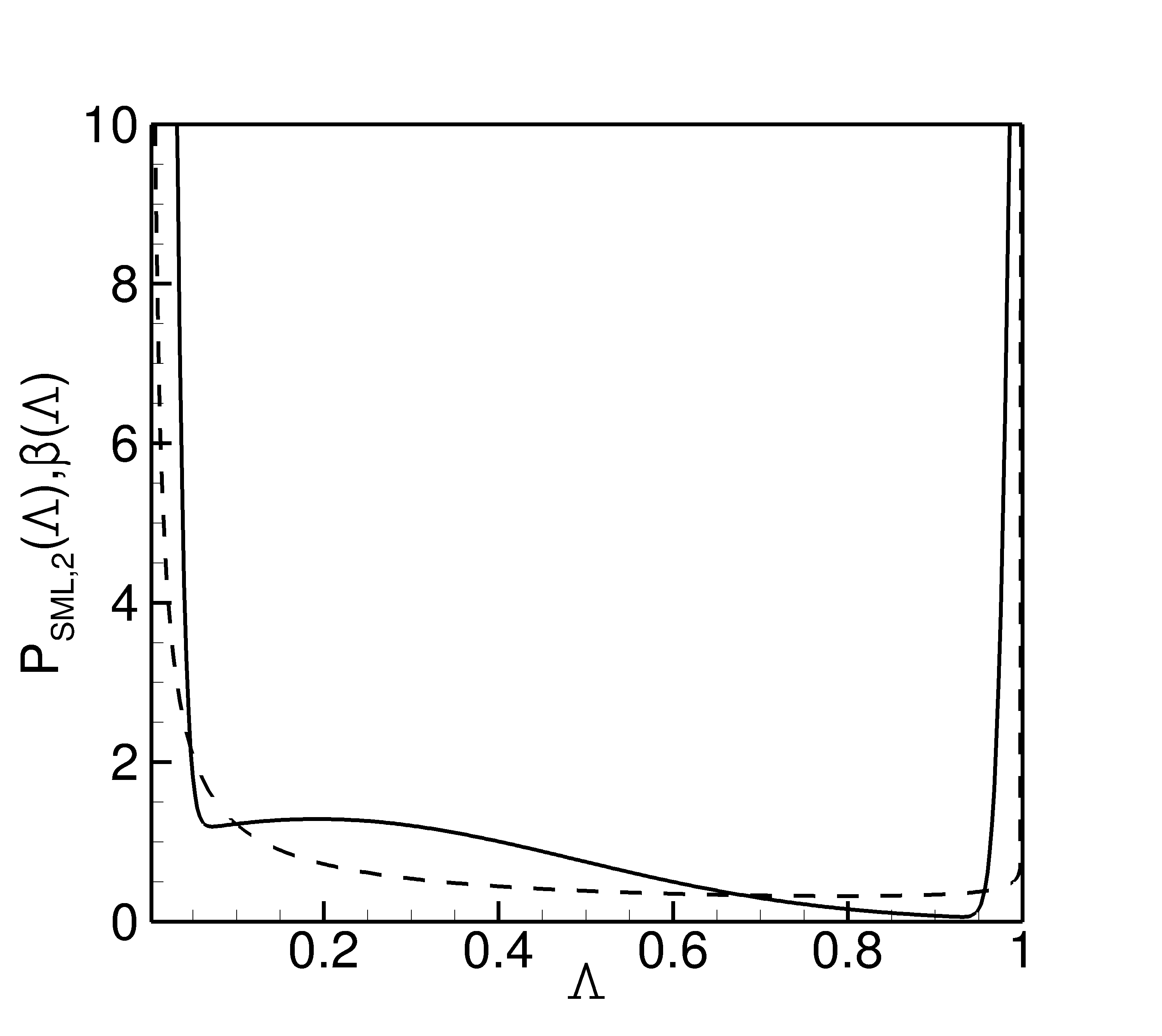}
\caption{Distribution of $Z$ and $\Lambda$ at the point $(x/d_{ref},y/d_{ref})=(20.85,1)$. The solid line is the $P_{SML,2}$ distribution and the dashed one the $\beta$-distribution.}
\label{x_20_85_y_1}
\end{center}
\end{figure}
In this section we analyse the influence of two widely used hypotheses for the FPV models, namely, the statistical independence of $Z$ and $\Lambda$ and the $\beta$-distribution PDF for $P(Z)$. To this purpose, two points have been selected in the flow field of the Sandia flame~E computed with model~D and the corresponding values of the mean and the variance of both $Z$ and $C$ have been used to evaluate the probability distributions. For the first point, with normalized coordinates $(x/d_{ref},y/d_{ref})=(0,1)$  (taken on the burner), the following values are obtained: $\widetilde Z=0.2700$, $\widetilde{Z''^2}=0.0034$, $\widetilde \Lambda=0.9618$ and $\widetilde{\Lambda''^2}=0.0009$. 
Note that the computational domain, shown in figure~\ref{grid}, includes a part of the burner so that the flow properties at the section $x/d_{ref}=0$ are computed whereas the boundary conditions are imposed at the inlet section at $x/d_{ref}=-1.5$.
The resulting PDFs are shown in figure~\ref{x_0_y_1}. It appears that the $\beta$-distribution and the SMLD distribution of $Z$ are in very good agreement. On the other hand, concerning the distribution of $\Lambda$, one can see that the two PDFs are quite different, providing different locations of the maxima and thus resulting in different evaluations of the mean values of the thermo-chemical quantities. It is worth noting that since $\beta(Z)$ and $P_{SML,2}(Z)$ are almost coincident, model~C and model~D differ only for the statistical independence hypothesis of $Z$ and $\Lambda$. 
This issue is further analysed in figure~\ref{scat}~(a), showing a scatter-plot corresponding to values for the mean and variance of $Z$ and $\Lambda$ computed at $(x/d_{ref},y/d_{ref})=(0,1)$. The joint~PDF is reported on both axes: the abscissa is evaluated using equation~\eqref{smld2} whereas the ordinate is computed by the Bayes theorem with the statistical independence hypothesis.  
The bisector is also reported as reference, so that, if the statistical independence hypothesis were correct, one would obtain a cloud of points aligned with the bisector. Figure~\ref{scat}(a) clearly shows that for Flame~E, the independence hypothesis is not appropriate in the region close to the burner.\\
Consider, now, the second point $(x/d_{ref},y/d_{ref})=(20.85,1)$, far away from the burner. Here, the value of mean and variance are: $\widetilde Z=0.3914$, $\widetilde{Z''^2}=0.0397$, $\widetilde \Lambda=0.2074$ and $\widetilde{\Lambda''^2}=0.0822$. At this location, the hypothesis that $Z$ is distributed according to a $\beta$-function results to be slightly less accurate, as shown in figure~\ref{x_20_85_y_1} (left panel).
The statistical independence hypothesis is slightly more appropriate than in the previous case; in fact one can see from figure~\ref{scat}~(b) that a non negligible part of the points are located near the bisector. Anyway, it still appears an incorrect hypothesis that should be abandoned in order to have an improvement in the combustion prediction.
\begin{figure}
\begin{center}
\includegraphics[scale=0.055]{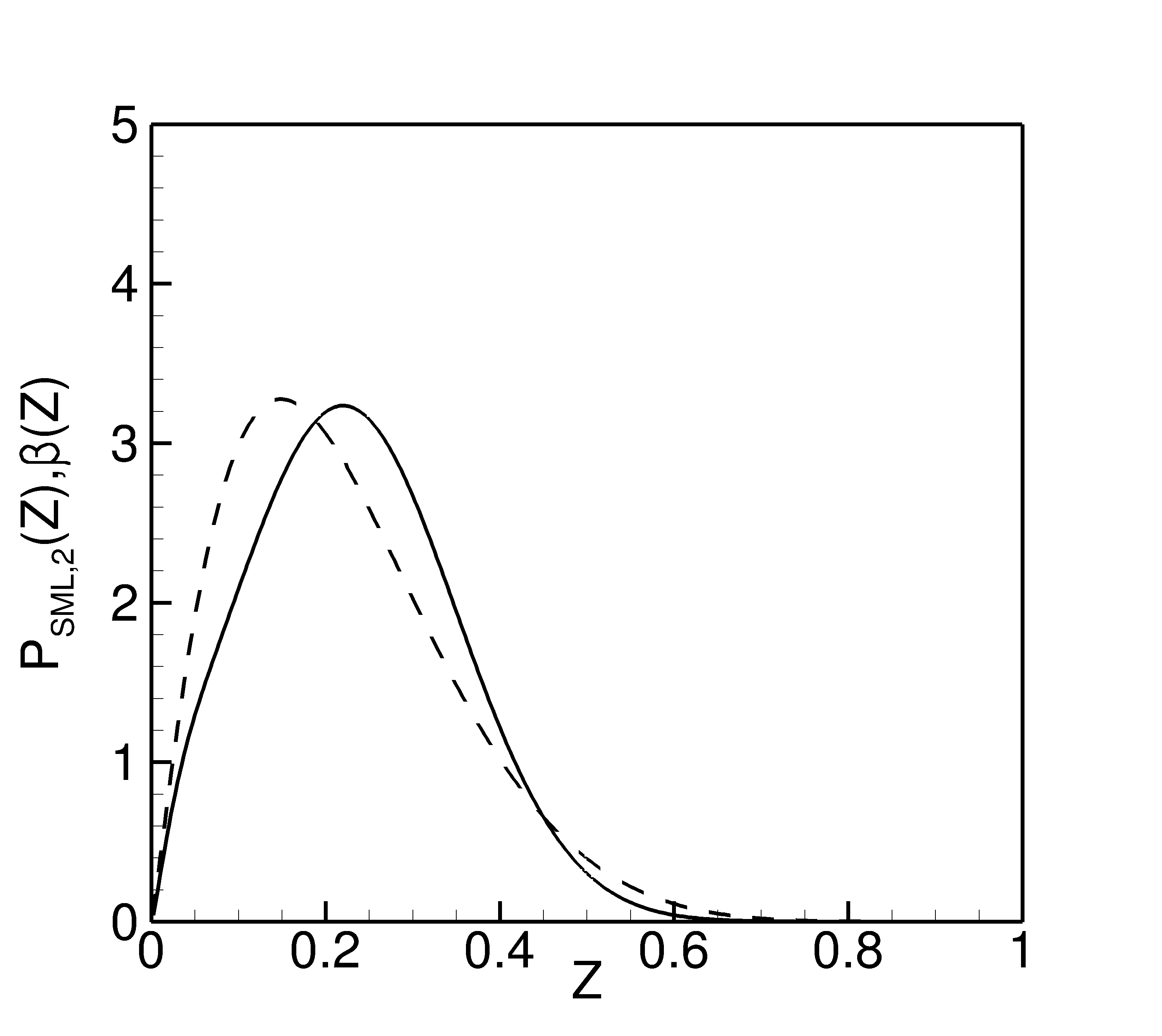}
\includegraphics[scale=0.055]{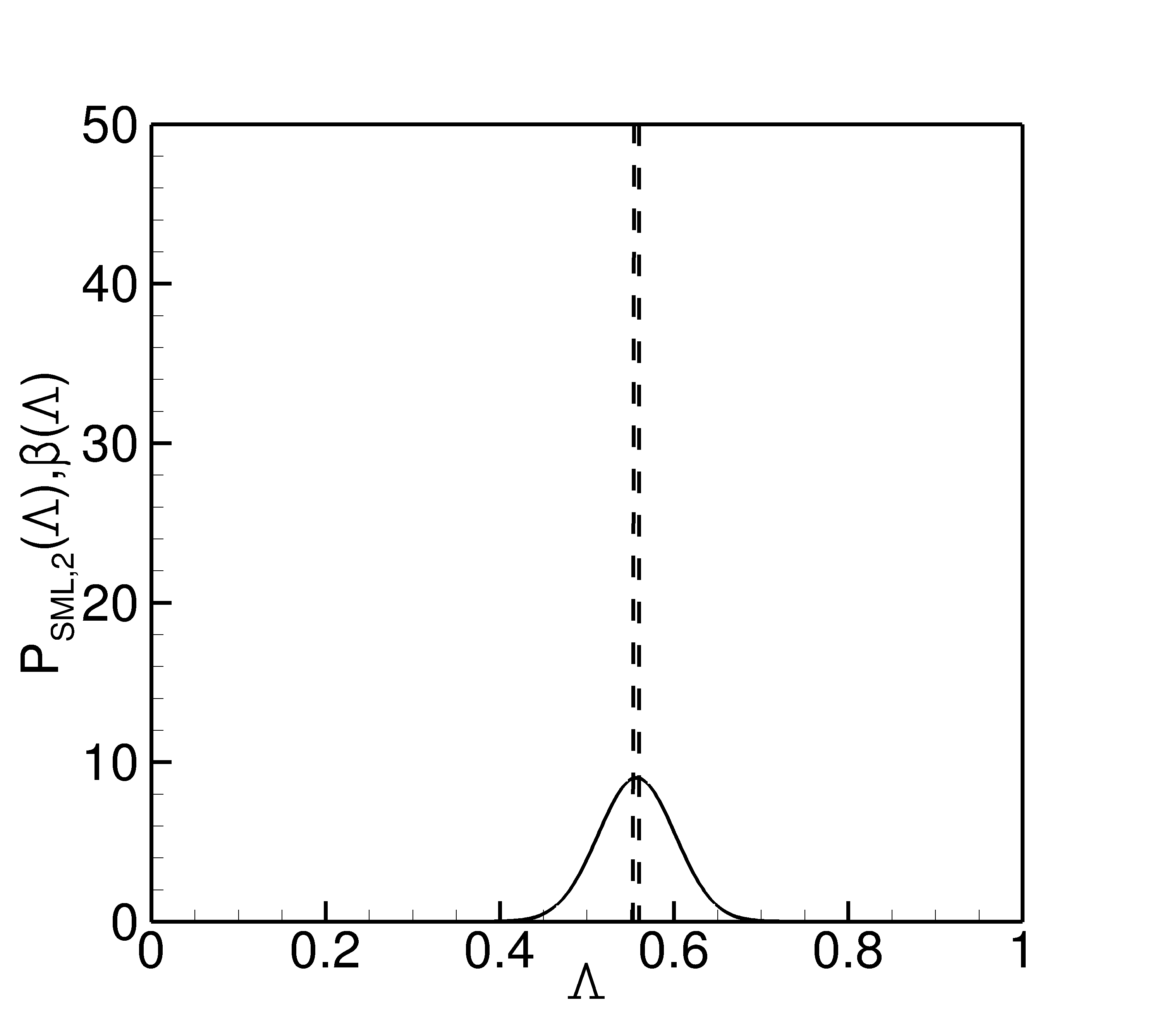}
\caption{Distribution of $Z$ and $\Lambda$ at the point $(x/d_{ref},y/d_{ref})=(1,1.21)$. The solid line is the $P_{SML,2}$ distribution and the dashed one the $\beta$-distribution.}
\label{x_1_y_121}
\end{center}
\end{figure}

Now let us analyse the suitability of the assumption of the $\beta$-distribution for $Z$ and for $\Lambda$ with an a-posteriori analysis. For this, three points are chosen at $x/d_{ref}=1$ (see the bottom-left panel of figure~\ref{rad}), where the difference between the results provided by model~A and model~D are found significant and the corresponding values of $\widetilde{Z}$, $\widetilde{Z''^2}$, $\widetilde{\Lambda}$, and $\widetilde{\Lambda''^2}$ are taken.
The first point, $y/d_{ref}=0.03$, is correctly evaluated by both models and the values of mean and variance obtained using model~D are: $\widetilde{Z}=1$, $\widetilde{Z''^2}=0$, $\widetilde{\Lambda}=6.055\times10^{-6}$, and $\widetilde{\Lambda''^2}=0$. 
It is clear that for these values $\beta(Z;\widetilde{Z},\widetilde{Z''^2})$ collapses in $\delta(Z-\widetilde{Z})$ and $\beta(\Lambda;\widetilde{\Lambda},\widetilde{\Lambda''^2})$ collapses in $\delta(\Lambda-\widetilde{\Lambda})$. In other words, the information related to the two values of the variance are lost, since the Dirac function is a one-parameter distribution depending only on the mean value. Using the $P_{SML,2}(Z;\widetilde{Z},\widetilde{Z''^2})$ and $P_{SML,2}(\Lambda;\widetilde{\Lambda},\widetilde{\Lambda''^2})$ in the same way, due to the low values of the variances, the distributions collapse in two Dirac distributions.
Very similar results are obtained for $y/d_{ref}=1.62$, where the values of mean and variance found by model~D are: $\widetilde{Z}=0$, $\widetilde{Z''^2}=0$, $\widetilde{\Lambda}=0.001$, and $\widetilde{\Lambda''^2}=6.449\times10^{-7}$.\\
The last point, $y/d_{ref}=1.21$, is very interesting because of the different results obtained with the two models. Here the values of mean and variance found by model~D are: $\widetilde{Z}=0.221$, $\widetilde{Z''^2}=0.016$, $\widetilde{\Lambda}=0.557$, and $\widetilde{\Lambda''^2}=0.002$. Figure~\ref{x_1_y_121} (left panel) shows that $\beta(Z;\widetilde{Z},\widetilde{Z''^2})$ and $P_{SML,2}(Z;\widetilde{Z},\widetilde{Z''^2})$ are very similar but $P_{SML,2}(\Lambda;\widetilde{\Lambda},\widetilde{\Lambda''^2})$ is more smooth than $\beta(\Lambda;\widetilde{\Lambda},\widetilde{\Lambda''^2})$ (right panel). In this case the differences between the PDFs are such that $Z_{Model D}=0.221$ and $Z_{Model A}=0.169$ whit a relative error of about $23.5\%$.
 
\section{Conclusions}
\label{conclusions}

This paper analyses the state-of-the-art constitutive hypotheses usually adopted for the definition of the presumed probability density function (PDF) in flamelet progress variable (FPV) models, discussing their adequateness and feasibility. Four combustion models are considered, based on different choices for the PDF of the mixture fraction and the progress parameter. Among these models, a more general model is proposed evaluating the most probable joint distribution of the mixture fraction and the progress parameter using a closure technique based on an analytical evaluation of the Lagrange's multiplier. The performance of the combustion models corresponding to different choices of the PDF are analysed by computing the Sandia flames D and E. The numerical results are also employed to study the validity of the hypothesis of statistical independence between the two variables. 
The results show that for RANS applications the proposed model provides a slight improvement in the prediction of the averaged thermo-chemical variables and a significant improvement in the evaluation of conditional quantities.
In particular, the temperature distributions and its peak values are closer to the experimental data. This is also confirmed by an a-posteriori analysis showing that the introduction of joint PDF for $Z$ and $\Lambda$ is effective. The proposed model relies on a more robust theoretical basis, with a substantially unchanged computational cost. Finally, we can reasonably expect that the differences found with the proposed model in evaluating the conditioned quantity should be less pronounced when using LES. However, since the computational overhead is marginal it is recommended to consider the joint PDF also for LES applications. Future work will be devoted to study the influence of the proposed FPV model in the LES framework.

\begin{acknowledgements}
This research has been supported by grant n. $PON02\_00576\_3333604$ INNOVHEAD. The authors are very grateful to Prof. M. D. de Tullio for his support in the software development. 
\end{acknowledgements}

\bibliographystyle{elsarticle-num.bst}      
\bibliography{biblio.bib}

\begin{thebibliography}{10}
\expandafter\ifx\csname url\endcsname\relax
  \def\url#1{\texttt{#1}}\fi
\expandafter\ifx\csname urlprefix\endcsname\relax\def\urlprefix{URL }\fi
\expandafter\ifx\csname href\endcsname\relax
  \def\href#1#2{#2} \def\path#1{#1}\fi

\bibitem{maas}
U.~Maas, S.~B. Pope, Simplifying chemical kinetics: intrinsic low-dimensional
  manifolds in composition space, Combust. Flame 88 (1992) 239--264.

\bibitem{piercemoin2004}
C.~D. Pierce, P.~Moin, Progress-variable approach for large-eddy simulation of
  non-premixed turbulent combustion, J. Fluid Mech. 504 (2004) 73--97.

\bibitem{laminarhydrogen}
O.~Gicquel, N.~Darabiha, D.~Thevenin, Laminar premixed hydrogen/air counterflow
  flame simulations using flame prolongation of {ILDM} with differential
  diffusion, Proc. Combust. Inst. 28 (2000) 1901--1908.

\bibitem{oijen}
J.~V. Oijen, L.~{De Goey}, Modelling of premixed laminar flames using
  flamelet-generated manifolds, Combust. Sci. Technol. 161 (2000) 113--137.

\bibitem{ihmeal2005}
M.~Ihme, C.~M. Cha, H.~Pitsch, Prediction of local extinction and re-ignition
  effects in non-premixed turbulent combustion using a flamelet/progress
  variable approach., Proc. Combust. Inst. 30 (2005) 793--800.

\bibitem{peters}
N.~Peters, Turbulent combustion, Cambridge University Press, 2000.

\bibitem{pierce}
C.~D. Pierce, Progress-variable approach for large-eddy simulation of turbulent
  combustion, {PhD Thesis}, Stanford University (2001).

\bibitem{ihmea}
M.~Ihme, H.~Pitsch, Prediction of extinction and re-ignition in non-premixed
  turbulent flames using a flamelet progress variable model. 1 {A} priori study
  and presumed {PDF}, Combust. Flame 155 (2008) 70--89.

\bibitem{ihmejcp2012}
M.~Ihme, L.~Shunn, J.~Zhang, Regularization of reaction progress variable for
  application to flamelet-based combustion models, J. {C}omput. {P}hys. 231
  (2012) 7715--7721.

\bibitem{cuenotCF2012}
A.~Najafi-Yazdi, B.~Cuenot, L.~Mongeau, Systematic definition of progress
  variables and intrinsically low-dimensional, flamelet generated manifolds for
  chemistry tabulation, Combust. Flame 159 (2012) 1197--1204.

\bibitem{vervischCF2013}
Y.-S. Niu, L.~Vervisch, P.~D. Tao, An optimization-based approach to detailed
  chemistry tabulation: Automated progress variable definition, Combust. Flame
  160 (2013) 776–785.

\bibitem{klimenko99}
A.~Klimenko, R.~Bilger,
  \href{http://www.scopus.com/inward/record.url?eid=2-s2.0-0032586559&partnerID=40&md5=4a2385f7be1501e6c2a5255bf9c6a98e}{Conditional
  moment closure for turbulent combustion}, Progress in Energy and Combustion
  Science 25~(6) (1999) 595--687, cited By (since 1996)425.
\newline\urlprefix\url{http://www.scopus.com/inward/record.url?eid=2-s2.0-0032586559&partnerID=40&md5=4a2385f7be1501e6c2a5255bf9c6a98e}

\bibitem{ihmeb}
M.~Ihme, H.~Pitsch, Prediction of extinction and re-ignition in non-premixed
  turbulent flames using a flamelet progress variable model. 2 {A}pplication in
  {LES} of {S}andia {F}lames {D} and {E}, Combust. Flame 155 (2008) 90--107.

\bibitem{DeMeesterCF2012}
R.~D. Meester, B.~Naud, B.~Merci, A priori investigation of {PDF}-modeling
  assumptions for a turbulent swirling bluff body flame ({‘SM1’}), Combust.
  Flame 159 (2012) 3353--3357.

\bibitem{AbrahamPoF2012}
S.~Mukhopadhyay, J.~Abraham, Evaluation of an unsteady flamelet progress
  variable model for autoignition and flame development in compositionally
  stratified mixtures, Phys. Fluids 24 (2012) 075115.

\bibitem{pope}
S.~B. Pope, {PDF} methods for turbulent reactive flows, Prog. Energy Combust.
  Sci. 11 (1985) 119--192.

\bibitem{pitsch98}
H.~Pitsch, N.~Peters, A consistent flamelet formulation of non-premixed
  combustion considering differential diffusion effects, Combust. Flame 114
  (1998) 317--332.

\bibitem{pitschchenpeters98}
H.~Pitsch, M.~Chen, N.~Peters, Unsteady flamelet modelling of turbulent
  hydrogen/air diffusion flames, Proceeding of the Combustion Institute 27
  (1998) 1057--1064.

\bibitem{kimwilliams93}
J.~S. Kim, F.~A. Williams, Structures of flow and mixture-fraction fields for
  counterflow diffusion flames with small stoichiometric mixture fractions,
  Journal of Applied Mathematics 53 (1993) 1551--1566.

\bibitem{peters84}
N.~Peters, Laminar diffusion flamelet models in non-premixed turbulent
  combustion, Prog. Energy Combust. Sci. 10 (1984) 319--339.

\bibitem{cook}
A.~W. Cook, J.~J. Riley, A subgrid model for equilibrium chemistry in turbulent
  flows, Phys. Fluids 6 (1994) 2868--2870.

\bibitem{jimenez}
J.~Jimenez, A.~Linan, M.~M. Rogers, F.~J. Higuera, A priori testing of subgrid
  models for chemically reacting non-premixed turbulent shear flows, J. Fluid
  Mech. 349 (1997) 149--171.

\bibitem{wall}
C.~Wall, B.~J. Boersma, P.~Moin, An evaluation of the assumed beta probability
  density function subgrid-scale model for large eddy simulation of
  non-premixed, turbulent combustion with heat release, Phys. Fluids 12 (2000)
  2522--2529.

\bibitem{cha_pitsch2002}
C.~M. Cha, H.~Pitsch, Higher-order conditional moment closure modelling of
  local extinction and reignition in turbulent combustion, Combust. Theory
  Model. 6 (2002) 425--437.

\bibitem{heinz}
S.~Heinz, Statistical mechanics of turbulent flows, Springer-Verlag, 2003.

\bibitem{shannon}
C.~H. Shannon, A mathematical theory of communication, Bell system technical
  journal 27 (1948) 379,423.

\bibitem{pope1979}
S.~B. Pope, A rational method of determining probability distributions in
  turbulent reacting flows, Journal of Non-Equilibrium Thermodynamics 4 (1979)
  309--842.

\bibitem{luigi}
L.~Cutrone, P.~{De Palma}, G.~Pascazio, M.~Napolitano, A {RANS}
  flamelet-progress-variable method for computing reacting flows of real-gas
  mixtures, Comput. Fluids 39 (2010) 485--498.

\bibitem{dsthesis}
D.~A. Schwer, {Numerical study of unsteadiness in non-reacting and reacting
  mixing layers}, {PhD Thesis}, The Pennsylvania State University (1999).

\bibitem{steger}
J.~L. Steger, R.~F. Warming, {Flux vector splitting of the inviscid gas-dynamic
  equations with applications to finite difference methods}, J. Comput. Phys.
  40 (1981) 263--293.

\bibitem{pulliam}
T.~H. Pulliam, D.~S. Chaussee, A diagonal form of an implicit factorization
  algorithm, J. Comput. Phys. 39 (1981) 347--363.

\bibitem{buelow97}
P.~E.~O. Buelow, D.~A. Schwer, J.-Z. Feng, C.~L. Merkle, D.~Choi, {A}
  preconditioned dual time diagonalized {ADI} scheme for unsteady computations,
  in: {A}IAA {P}roceedings, 1997.

\bibitem{menter}
F.~Menter, C.~Rumsey, Assessment of two-equation turbulence models for
  transonic flows, in: 25$^{th}$ AIAA Fluid Dynamics Conference, AIAA, Colorado
  Springs, CO, 1994.

\bibitem{sandia}
Sandia National Laboratories, TNF Workshop,
  \texttt{http://www.ca.sandia.gov/TNF}.

\bibitem{flamemaster}
H.~Pitsch, Flamemaster v3.3. a c++ computer program for 0d combustion and 1d
  laminar flame calculationsAvailable at \texttt{http://www.stanford.edu/$\sim$
  hpitsch}.

\bibitem{barlow98}
R.~Barlow, J.~Frank, Effects of turbulence on species mass fraction in
  methane/air jet flames, Proc. Combust. Inst. 27 (1998) 1087--1095.

\bibitem{grimech30}
G.~P. Smith, D.~M. Golden, M.~Frenklach, N.~W. Moriarty, B.~Eiteneer,
  M.~Goldenberg, C.~T. Bowman, R.~K. Hanson, S.~Song, W.~C. Gardiner, V.~V.
  Lissianski, Z.~Qin, 2000, \texttt{http://www.me.berkeley.edu/gri\_mech/}.

\bibitem{ochoa2012}
J.~Ochoa, A.~Sánchez-Insa, N.~Fueyo, Subgrid linear eddy mixing and combustion
  modelling of a turbulent nonpremixed piloted jet flame, Flow, Turbulence and
  Combustion 89~(2) (2012) 295--309.

\bibitem{zoller2011}
B.~T. Zoller, J.~M. Allegrini, U.~Maas, P.~Jenny, {PDF} model for {NO}
  calculations with radiation and consistent {NO-NO2} chemistry in non-premixed
  turbulent flames, Combustion and Flame 158~(8) (2011) 1591--1601.

\bibitem{kemenov2011}
K.~A. Kemenov, S.~B. Pope, Molecular diffusion effects in {LES} of a piloted
  methane-air flame, Combustion and Flame 158~(2) (2011) 240--254.

\bibitem{ferraris2008}
S.~Ferraris, J.~Wen, {LES} of the sandia flame {D} using laminar flamelet
  decomposition for conditional source-term estimation, Flow, Turbulence and
  Combustion 81~(4) (2008) 609--639.

\bibitem{juddoo2011}
M.~Juddoo, A.~R. Masri, S.~B. Pope, Turbulent piloted partially-premixed flames
  with varying levels of {O2/N2}: Stability limits and {PDF} calculations,
  Combustion Theory and Modelling 15~(6) (2011) 773--793.
\newblock \href {http://dx.doi.org/10.1080/13647830.2011.563867}
  {\path{doi:10.1080/13647830.2011.563867}}.

\end{thebibliography}
\end{document}